\documentclass[journal]{IEEEtran}
\usepackage{caption}
\usepackage{algorithmic}
\usepackage[linesnumbered,ruled]{algorithm2e}
\usepackage{graphics,epstopdf}
\usepackage{balance}
\usepackage{listings}
\usepackage{comment}
\usepackage{enumitem}
\usepackage{fixmath}
\usepackage{cite}
\usepackage{amsmath,amssymb,amsfonts}
\usepackage{graphicx}
\usepackage{textcomp}
\usepackage{tabularx}
\usepackage{booktabs}
\usepackage{longtable}
\usepackage{multirow}
\usepackage{colortbl}
\usepackage{hhline}
\usepackage{tabularx,colortbl}
\usepackage{stmaryrd}
\usepackage{cite}
\usepackage{enumitem}
\usepackage{soul}
\usepackage{bm} 
\usepackage{mathtools} 
\usepackage{mathrsfs} 
\usepackage{multirow}
\usepackage{array}
\usepackage{colortbl} 
\usepackage{multirow, booktabs}
\usepackage{hhline, boldline, makecell}

\usepackage{cellspace}
\usepackage[table,xcdraw]{xcolor}

\ifCLASSINFOpdf

\else

\fi

\begin{document}

\title{Detection of False-Reading Attacks in the AMI Net-Metering System}

\author{\IEEEauthorblockN{Mahmoud M. Badr, Mohamed I. Ibrahem,
Mohamed Mahmoud,~\IEEEmembership{Senior~Member,~IEEE,}\\
Mostafa M. Fouda,~\IEEEmembership{Senior~Member,~IEEE,} and
Waleed Alasmary,~\IEEEmembership{Senior~Member,~IEEE,}  \\}
\thanks{Corresponding author: Mahmoud M. Badr.}
\thanks{M. M. Badr, M. I. Ibrahem, and M. Mahmoud are with the Department of Electrical and Computer Engineering, Tennessee Tech. University, Cookeville, TN 38505 USA (e-mail: mmbadr42@tntech.edu; miibrahem42@tntech.edu; mmahmoud@tntech.edu).}
\thanks{M. M. Fouda is with the Department of Electrical and Computer Engineering, College of Science and Engineering, Idaho State University, Pocatello, ID 83209, USA (e-mail: mfouda@ieee.org).}
\thanks{W. Alasmary is with the Department of Computer Engineering, Umm Al-Qura University, Saudi Arabia (e-mail: wsasmary@uqu.edu.sa).}
\thanks{
Copyright (c) 20xx IEEE. Personal use of this material is permitted. However, permission to use this material for any other purposes must be obtained from the IEEE by sending a request to pubs-permissions@ieee.org.
} 
}

\maketitle

\markboth{Badr \MakeLowercase{\textit{et al.}}: Detection of False-Reading Attacks in the AMI Net-Metering System}%
{}
\IEEEpeerreviewmaketitle
\begin{abstract}
In the advanced metering infrastructure (AMI) network of the smart power grid, smart meters (SMs) are installed at the customers' premises to report their fine-grained power consumption readings to the utility for billing and load monitoring purposes. Moreover, to create a clean power system, customers install solar panels on their rooftops to generate power and sell it to the utility. However, malicious customers may compromise their SMs to report false readings to achieve financial gains illegally. Reporting false readings not only causes hefty financial losses to the utility but may also degrade the grid performance because the reported readings are used for energy management. This paper is the first work that investigates this problem in the net-metering system, in which one SM is used to report the difference between the power consumed and the power generated. First, we prepare a benign dataset for the net-metering system by processing a real power consumption and generation dataset. Then, we propose a new set of attacks tailored for the net-metering system to create malicious dataset. After that, we analyze the data and we found time correlations between the net meter readings and correlations between the readings and relevant data obtained from trustworthy sources such as the solar irradiance and temperature. Based on the data analysis, we propose a general multi-data-source deep hybrid learning-based detector to identify the false-reading attacks. Our detector is trained on net meter readings of all customers besides data from the trustworthy sources to enhance the detector performance by learning the correlations between them. The rationale here is that although an attacker can report false readings, he cannot manipulate the solar irradiance and temperature values because they are beyond his control. Extensive experiments have been conducted, and the results indicate that our detector can identify the false-reading attacks with high detection rate and low false alarm.

\end{abstract}

\section{Introduction}
\label{intro}
Smart power grid is a new vision that aims to upgrade the traditional power grid to create a clean, efficient and resilient system. Advanced metering infrastructure (AMI) is one of the main components of the smart power grid, where smart meters (SMs) are installed at the customers' premises to periodically report fine-grained power consumption readings to the utility for billing and load monitoring purposes \cite{jokar}. Moreover, to create a clean system, the smart power grid aims to generate more electricity from renewable resources, e.g., solar panels, to reduce the emissions of greenhouse gases \cite{ismail2018}. To do that, solar panels are installed on the rooftops of the customers to generate power and sell it to the utility.
Therefore, in the smart grid, some houses may have renewable energy generators and other houses do not generate power. In the latter case, there is only one metering system adopted by the utilities, called the \textit{consumption metering} system, where each house is equipped with one SM to measure the power consumption readings and send them to the utility. While in case that houses generate power, there are two metering systems adopted by the utilities, namely, the \textit{feed-in tariff} (\textit{FIT}) and the \textit{net-metering}, to enable the customers to sell their generated power \cite{ismail2018, ismail2020}. 
 
In the FIT system, the tariff of the power consumed by the customers is different from the tariff of the power generated by them \cite{ismail2020,ismail2018}. In this case, the customer's home is equipped with two SMs; a consumption meter which is used for reporting the power consumption readings and a generation meter which is used for reporting the power generation readings. 
On the other hand, in the net-metering system, the tariff of the power sold by the customers is similar to the tariff of the power consumed by them \cite{netwhere}. Hence, the excess generated power can be injected directly to the grid, and thus, the customer does not need to purchase an expensive solar battery. This can significantly reduce the cost of the solar generation system, which motivates the customers to install it. Also, in the net-metering system, only one SM, called net meter, is used to report readings which represent the difference between the power consumed and the power generated by the customer in a small time period \cite{ismail2020}. Therefore, the reading is positive if the consumed power is more than the generated power, and it is negative if the generated power is more than the consumed power. The customer is charged for the positive readings and rewarded for the negative readings. 
Given the advantages of the net-metering system, it is currently adopted in many countries worldwide including USA, Italy, and Brazil \cite{netwhere}.

In these metering systems, malicious customers can report false readings to the utility to make profit illegally. Specifically, in the consumption metering system, malicious customers can report lower readings to reduce their bills \cite{jokar}, and in the FIT system, malicious customers can report higher generation readings to achieve higher financial profit \cite{ismail2018}. Moreover, malicious customers in the net metering can report lower readings when the consumed power is more than the generated power and report higher readings when the generated power is more than the consumed power.

Reporting false consumption readings for electricity theft, which is a contemporary problem that faces the utilities all over the world, causes hefty financial losses. According to \cite{news}, the world annual losses due to electricity theft were estimated by 89.3 billion dollars. For instance, the United States and Puerto Rico lose about 6 billion and 400 million dollars every year, respectively \cite{jokar,ismail2018}. Moreover, the false readings may degrade the grid performance because they are used to make decisions regarding energy management \cite{endesa1}. 
To detect the false-reading attacks (i.e, false reported readings by malicious customers) in the AMI network, various solutions have been proposed in the literature \cite{jokar,seraj2014smart,endesa1,kenz2016,course,espain,ismail2018,ismail2020}. However, all the existing works study only the consumption metering \cite{jokar,seraj2014smart,endesa1,kenz2016,course,espain} and FIT systems \cite{ismail2018, ismail2020}, and \textit{none of the existing works have studied the problem in the net-metering system}. 

Detection of false-reading attacks in the net-metering system is different from the other metering systems for the following reasons. In case of the consumption metering system, the detector can be trained on the consumption pattern of the customer, which depends on his lifestyle, to detect the false readings. Similarly, in the FIT system, the detector can be trained on the generation pattern of the customer's solar panels to detect the false readings. However, the problem is more complicated in case of the net-metering system because the net meter readings depend on the lifestyle, the solar irradiance, and the generation capacity of the solar panels, i.e., the readings simultaneously include consumption and generation patterns. This means that a new detection approach that considers both the consumption and the generation patterns is needed to be able to detect the false-reading attacks in the net-metering system.
Furthermore, new attacks tailored for the net-metering system should be investigated for the following reason. The attacks against the consumption metering system target reducing the readings while trying to mimic the consumption pattern, and the attacks against the FIT system target increasing the generation readings while trying to mimic the generation pattern. However, in the net-metering system, the attacker needs to consider both the consumption and generation patterns in computing the false readings while achieving financial gains.

In this paper, we investigate the detection of false-reading attacks in the net-metering system using deep learning. Our methodology consists of four steps: dataset preparation, data analysis, detector design, and performance evaluation. To prepare our dataset, the real power consumption and generation dataset of Ausgrid \cite{ausgrid} is used to derive  benign samples of true net meter readings. Then, a set of attacks that mimic the behavior of malicious customers is proposed to create malicious samples of false readings. The dataset is extended by including weather information collected from SOLCAST website \cite{solcast}. After that, the data is analyzed, and time correlations are found between the consecutive readings of the benign samples. Moreover, correlations are found between the true net meter readings and the relevant data from trustworthy sources such as the solar irradiance and temperature. 
Based on the data analysis, a general multi-data-source hybrid deep learning-based detector is proposed to identify the false-reading attacks. 

Our general detector can be applied for all customers, and it has a hybrid architecture that includes a convolutional neural network (CNN) and a gated recurrent unit neural network (GRU). This hybrid architecture is used so that the CNN layers extract the features from the input net meter readings while the GRU layers capture the correlation between the extracted features. Moreover, our detector is trained on the net meter readings besides the relevant data from trustworthy sources, such as the solar irradiance and temperature, to enhance the detection performance by learning the correlations between them. The rationale here is that although an attacker can report false readings, he cannot manipulate the solar irradiance and temperature values because they are beyond his control. Thus, the true data from the trustworthy sources can help the detector to identify the false-reading attacks. The simulation results of our experiments indicate that our detector can accurately detect the false reading attacks and achieve a higher performance than a single-data-source detector trained only on the net meter readings.

To the best of our knowledge, this is the first work that investigates the detection of false-reading attacks in the net-metering system, and our main contributions can be summarized as follows.
\begin{itemize}
\item We prepare a benign dataset for the net-metering system by processing the Ausgrid dataset \cite{ausgrid} and exploiting the weather information available on SOLCAST website \cite{solcast}. We also propose a set of attacks tailored for the net metering system to mimic the behavior of malicious customers to create a malicious dataset.

\item We analyzed the dataset and found time correlations between the net meter readings and correlations between the readings and relevant data obtained from trustworthy sources. Based on this data analysis, we propose a multi-data-source hybrid deep learning-based detector to identify false-reading attacks in the net metering system. Our detector uses the net meter readings with relevant data obtained from trustworthy sources to enhance the performance by learning the correlations between the readings and the other data. These data include the solar irradiance, the temperature, the solar panel capacity, the day, and the season.

\item We conduct extensive experiments to evaluate the performance of our multi-data-source detector, and the results indicate that our detector can accurately detect the false-reading attacks. Furthermore, our detector  achieves higher performance (i.e., higher detection rate and lower false alarm) compared to a single-data-source detector trained only on the net meter readings.

\end{itemize}

The rest of the paper is organized as follows. In Section \ref{relate}, we discuss the existing works in the literature that address detecting false-reading attacks in the AMI network. Then, the network and threat models are discussed in Section \ref{models}. Section \ref{sec:Preliminaries} presents the preliminaries used in our work.
The dataset created for training our detector is presented in Section \ref{prepare}. Our detector designed to identify false-reading attacks is presented in Section \ref{chap:design}. Next, performance evaluation of our detector is discussed in Section \ref{performanceEvaluation}. Finally, the paper is concluded in Section \ref{conclusion}.

\section{Related Work}
\label{relate}
In this section, we discuss the research works that address detecting false-reading attacks in the AMI network of the smart power grid using machine learning approaches. 
These works either consider the consumption metering system \cite{jokar,seraj2014smart,endesa1,kenz2016,course,espain} or the FIT system \cite{ismail2018, ismail2020}. Then, we will discuss the limitations and research gap.

\subsection{The Consumption Metering System}
Various solutions have been proposed in the literature to detect false-reading attacks in the consumption metering system. While some of these solutions use shallow detectors \cite{jokar, seraj2014smart, endesa1}, other solutions use deep learning-based detectors \cite{kenz2016, course, espain}.

\subsubsection{Shallow Detectors}
Jokar \textit{et. al.} \cite{jokar} and Ford \textit{et. al.} \cite{seraj2014smart} have proposed false-reading attacks detector using the Irish dataset \cite{irish} that contains benign samples of real consumption readings. A set of attacks have been proposed in \cite{jokar} to create synthetic malicious samples. Then, two  support vector machine (SVM)-based detectors are used for each customer; the first detector is a one-class SVM trained only on the benign samples, and the other is a multi-class SVM trained on both the benign and malicious samples. The results in \cite{jokar} indicate that the multi-class SVM gives superior performance than the one-class SVM.
Unlike \cite{jokar} that trains customer-specific detectors, i.e., a detector for each customer, the detector in \cite{seraj2014smart} is general so that it can be applied for all customers. The detector is based on an artificial neural network (NN) with single hidden layer. It uses the historical consumption readings of customers to predict the future consumption values which are compared with the reported consumption values using the root mean squared error. If this error exceeds a threshold, the customer is assumed malicious, otherwise he is honest. 

Buzau \textit{et. al.}\cite{endesa1} have trained a general detector using the dataset of Endesa \cite{endesa1, espain}, the largest electricity utility in Spain, that contains both benign and malicious samples. To enhance the detection of false-reading attacks, some information in addition to the consumption readings are taken into account such as the geographical locations of the customers and the technological characteristics of the SMs. The detector in \cite{endesa1} uses extreme gradient boosted trees (XGBoost), and the results indicate that the XGBoost-based detector outperforms the other detectors based on SVM, logistic regression, and K-nearest neighbors.

\subsubsection{Deep Learning-Based Detectors}
There are detectors that use deep learning to identify the false-reading attacks \cite{kenz2016, course, espain}.
Unlike shallow detectors which need feature extraction techniques to successfully capture the behavior of the input data, the deep learning-based detectors can automatically extract these features through their deep layers. 
A synthetic dataset is used in \cite{kenz2016} to train different types of deep learning-based general detectors using CNN, long short-term memory network (LSTM), and Stacked Autoencoder as well as shallow detectors using decision tree (DT), random forest (RF), and shallow NN. The results indicate that deep learning-based detectors outperform the shallow detectors, while the CNN-based detector achieves the highest performance among all detectors.

Zheng \textit{et. al.} \cite{course}  have trained a general detector using the state grid corporation of China (SGCC) dataset \cite{SGCC} that contains both benign and malicious samples to detect false-reading attacks.
The detector uses a deep learning architecture which includes both multi-layer perceptron (MLP) and CNN, and the results indicate that the proposed detector outperforms shallow, MLP-based, and CNN-based detectors. 
Buzau \textit{et. al.} \cite{espain} have used the Endesa dataset to train a general detector. The proposed detector uses a deep learning architecture which includes an LSTM module and an MLP module. Sequential data (e.g., daily average power consumption) and non-sequential data (e.g., the SM model, the location, and the contracted power) have been used for detecting the false-reading attacks. The results indicate that the  detection accuracy of the proposed detector is better than that of \cite{course}.

\begin{figure*}
    \centering
    \includegraphics[scale=0.48]{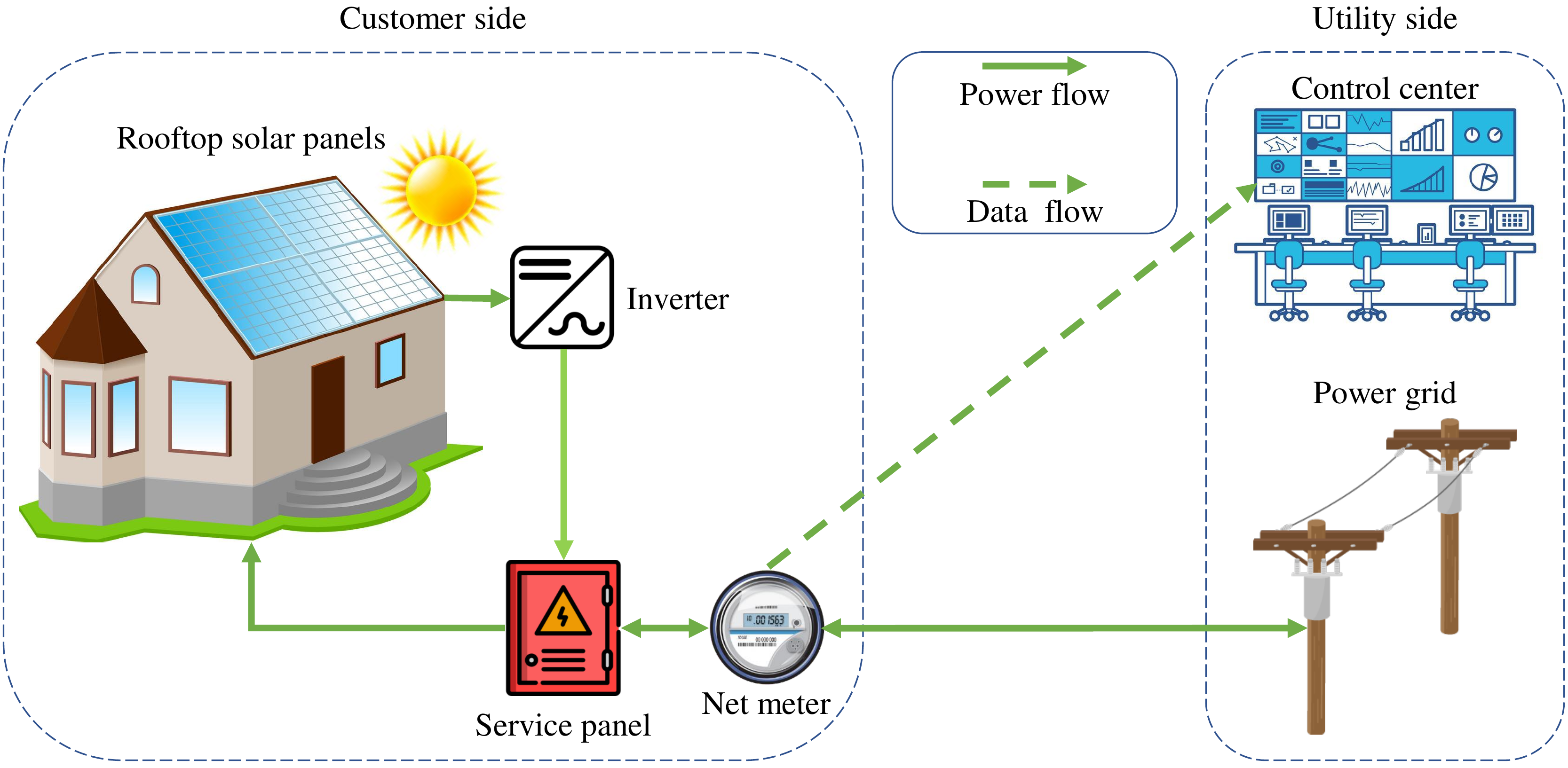}
    \caption{The net-metering system model.}
    \label{overview}
\end{figure*}


\subsection{The FIT System}
A few works in the literature have addressed detecting false-reading attacks in the FIT system \cite{ismail2018, ismail2020}. Krishna \textit{et. al.} \cite{ismail2018} have proposed different approaches to design customer-specific anomaly detectors based on the auto-regressive integrated moving average (ARIMA) and the Kullback-Leibler divergence (KLD) to detect false-reading attacks in the generation domain of the FIT system. The proposed detectors have been trained only on the benign samples of various datasets including the Ausgrid dataset \cite{ausgrid}. Krishna \textit{et. al.} \cite{ismail2018} have proposed a set of attacks, that can maximize the financial profit of the attacker by reporting false readings, to evaluate the performance of the proposed detectors. The lower the financial profit of the attacker, the more robust the detector against false-reading attacks.
Unlike \cite{ismail2018} that trains customer-specific detectors, Ismail \textit{et. al.} \cite{ismail2020} have trained a general detector using a synthetic dataset. A set of attacks has been proposed to generate malicious samples from the benign samples of the synthetic dataset. Unlike the detectors in \cite{ismail2018} which are trained on benign samples only, the detector in \cite{ismail2020} is trained on both benign and malicious samples. The proposed detector has a deep learning architecture, and the results indicate that the proposed detector achieves a higher performance than the detectors in \cite{ismail2018}.



 

\subsection{Limitations and Research Gap}
As discussed in the previous two subsections, all the existing works consider only the consumption metering and FIT systems, and detecting false-reading attacks in the net-metering system has not been addressed. The net-metering system is a practical system that is currently adopted in many countries including USA, Italy, and Brazil \cite{netwhere}. Thus, this paper tries to fill the research gap by investigating a deep learning-based detector to detect false-reading attacks in the net-metering system.

Detection of false-reading attacks in the net-metering system is different from the other metering systems for the following reasons. In case of the consumption metering system, the detector can be trained on the consumption pattern of the customer, which depends on his lifestyle, to detect the false readings. Similarly, in the FIT system, the detector can be trained on the generation pattern of the customer's solar panels to detect the false readings. However, the problem is more complicated in case of the net-metering system because the net meter readings depend on the lifestyle, the solar irradiance, and the generation capacity of the solar panels, i.e., the readings simultaneously include consumption and generation patterns. This means that a new detection approach that considers both the consumption and the generation patterns is needed to be able to detect the false-reading attacks in the net-metering system. What makes the problem more difficult is that to devise a general detector that can be applied for all customers, it needs to be trained on data from different customers who have different lifestyle, type of solar panels, and generation capacity.

Furthermore, new attacks tailored for the net-metering system should be investigated for the following reasons. The attacks against the consumption metering system target reducing the readings while trying to mimic the consumption pattern, and the attacks against the FIT system target increasing the generation readings while trying to mimic the generation pattern. However, in the net-metering system, the attacker needs to consider both the consumption and generation patterns in computing the false readings while achieving financial gains.
Moreover, the majority of the research works in the consumption metering system have proposed simple attacks such as reporting zero readings \cite{jokar, seraj2014smart, endesa1,kenz2016, course,espain} or continuously reporting the same reading during the successive periods \cite{jokar}. Also, the research works in the FIT system have proposed simple attacks such as reporting readings higher than the generation capacity of the solar panels \cite{ismail2018, ismail2020} or reporting generation readings higher than zero after the sunset \cite{ismail2020}. Such attacks can be easily detected even without using machine learning techniques. However, in this paper, we avoid these limitations by proposing more sophisticated attacks that consider the generation capacity of the solar panels and try to mimic the consumption and generation patterns.

\section{System and Threat Models}\label{models}
In this section, we discuss the system and threat models considered in this paper.
\subsection{System Model}
Fig. \ref{overview} shows the different entities in our net-metering system and the interactions between them. At the customer side, there are solar panels installed on the rooftop of the customer's house. These solar panels convert the energy collected from the sun to  direct current (DC) electricity which cannot be used directly to power the house's appliances or injected to the grid. Therefore, an inverter is used to convert the DC electricity to alternating current (AC) electricity. Then, the AC electricity flows from the inverter to the main service panel that controls feeding the electricity to the house's appliances. 
In the net-metering system, the customer's solar generation system is connected to the utility's power grid as shown in Fig. \ref{overview} so that the excess power generated can be injected directly to the grid through the net meter, and thus, the customer does not need to purchase an expensive solar battery. On the other hand, if the generated power is insufficient, the customer can satisfy his consumption needs by drawing electricity from the power grid through the net meter. Thus, the net meter acts as an interface between the customer and the utility, and it is a bidirectional meter that allows the electricity flow in either direction. 

The net meter periodically records readings, which represent the difference between the power consumed by the house appliances and the power generated by the solar panels.
Therefore, the reading is positive if the consumed power is more than the generated power, the reading is negative if the generated power is more than the consumed power, and the reading is zero if the consumed power is similar to the generated power.
These fine-grained readings are communicated to the utility's control center through the AMI network via wired communications such as power line communication or wireless communication such as cellular communication \cite{wang2011survey}. These readings are used by the utility for billing purposes and for demand side management (i.e, achieving a balance between the energy demand and supply).

\subsection{Threat Model}
Given the high installation costs of the solar generation system, customers may be keen to get as high profits as they can from the system even by illegal ways to shorten the time taken to compensate the paid costs. In doing so, malicious customers may launch false-reading attacks by compromising their net meters to report false readings to the utility to achieve financial gains illegally. Specifically, malicious customers can report lower readings when the consumed power is more than the generated power (in case of positive readings) and report higher readings when the generated power is more than the consumed power (in case of negative readings). 

SMs can be compromised to report false readings by programming a malicious firmware and installing it in the SM that is accessible through the ANSI optical port \cite{ismail2018}. This port is usually secured via weak passwords and there are some tools, such as Terminator, that are used to launch brute force attack to guess the passwords and gain access to the SM \cite{ismail2018}.
Moreover, recent studies have shown that the AMI communication networks have vulnerabilities, which can be exploited by malicious customers to launch the false-reading attacks \cite{8039510}.
The false reported readings not only cause financial losses to the utility but also can result in wrong decisions regarding energy management. Thus, in this paper, we propose a deep learning-based detector to analyze the readings reported by the customers' net meters to detect the false-reading attacks.

\section{Preliminaries} \label{sec:Preliminaries} 
We present, in this section, a brief description of the deep learning approaches and the popular activation functions (AFs) that will be used in our false-reading attacks detector.


\subsection{Deep Learning}
Deep learning model is a neural network which has multiple hidden layers. Generally, the neural network composes of input, output, and hidden layers~\cite{zheng2018wide}. 
Deep learning is a promising technique to many applications like face recognition~\cite{6549322} and using voice for age identification~\cite{8404322} because of its high accuracy. 
In this paper, we use different deep learning models to detect false-reading attacks in the net-metering system. This is a classification problem which needs one of the supervised learning methods that use a labeled dataset to train a model. There are various types of supervised learning models including the MLP~\cite{10.5555/1213811}, CNN~\cite{lecun1995convolutional}, and recurrent neural network (RNN)~\cite{ha2018recurrent}.

The aim of the training process of a model is to obtain good values for all the model weights and biases. This can be done by defining an objective function and using an optimizer and labeled data samples. First, the input data goes from the first layer in the model through the intermediate layers for a predefined number of iterations. Then, the model's weights and biases are updated in each iteration in the direction of minimizing the objective function $\Theta$ using feed-forward and back-propagation~\cite{6745416}. 
The most widely used objective function in the classification problems is categorical cross-entropy $C(y,\hat{y})$, and it measures the loss between the true distribution $y$ and the predicted distribution $\hat{y}$, for $N$ classes as follows: 

\begin{equation}
    \begin{aligned}
        C(y,\hat{y}) = \underset{\Theta}{\min} (-\sum _{c\,=\,1}^{N} y(c)\ log(\hat{y}(c)))\label{eq:cross-entropy}
    \end{aligned}
\end{equation}


In our paper, the MLP, CNN, and GRU are used to train the false-reading attacks detector.

\subsection {Feed-Forward Neural Network (FFN)}
\begin{figure}[t]
\centering
\includegraphics[width=\columnwidth]{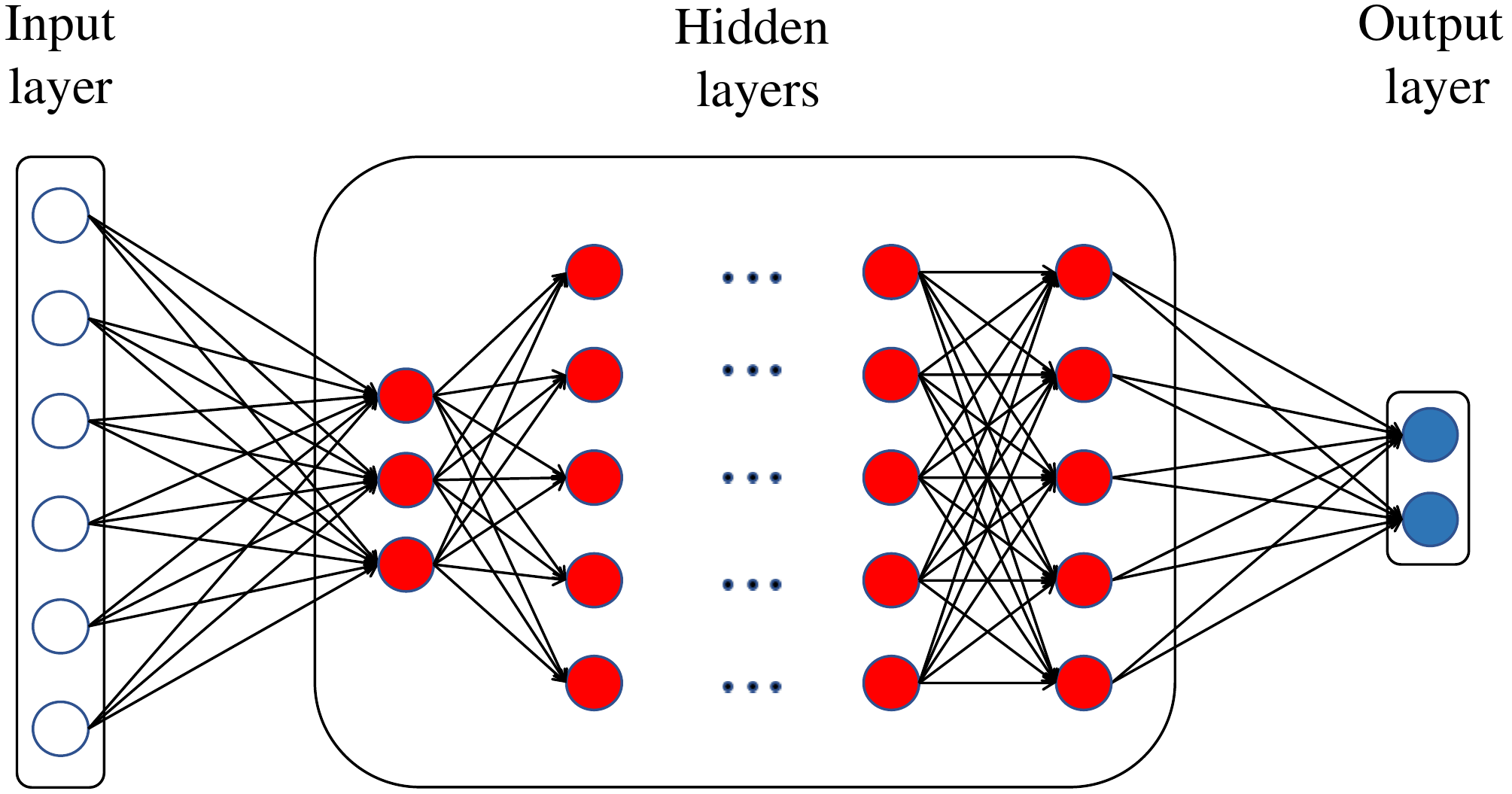}
\caption{Typical architecture of a feed-forward neural network (FFN).} \label{fig:ffn_arch}
\end{figure}

FFN is also called MLP~\cite{10.5555/1213811}, and it consists of three types of layers as can be seen in Fig. \ref{fig:ffn_arch} as follows.
\begin{itemize}
    \item \textit{Input Layer}: This is the first layer of an FFN, and it passes the input data to the following layers through nodes, called neurons.

    \item \textit{Output Layer}: This is the last layer that is responsible for determining the output (or classification) of the model. 
    
    \item \textit{Hidden Layers}: These are the intermediate layers between the input and output layers. Each hidden layer composes of a number of neurons, where each neuron uses an activation function to transform its input values into the output value of that neuron. Every neuron is fully connected to the neurons of the previous layer through a number of connections. 
    
\end{itemize}


\begin{figure}[t]
\centering
\includegraphics[width=\columnwidth]{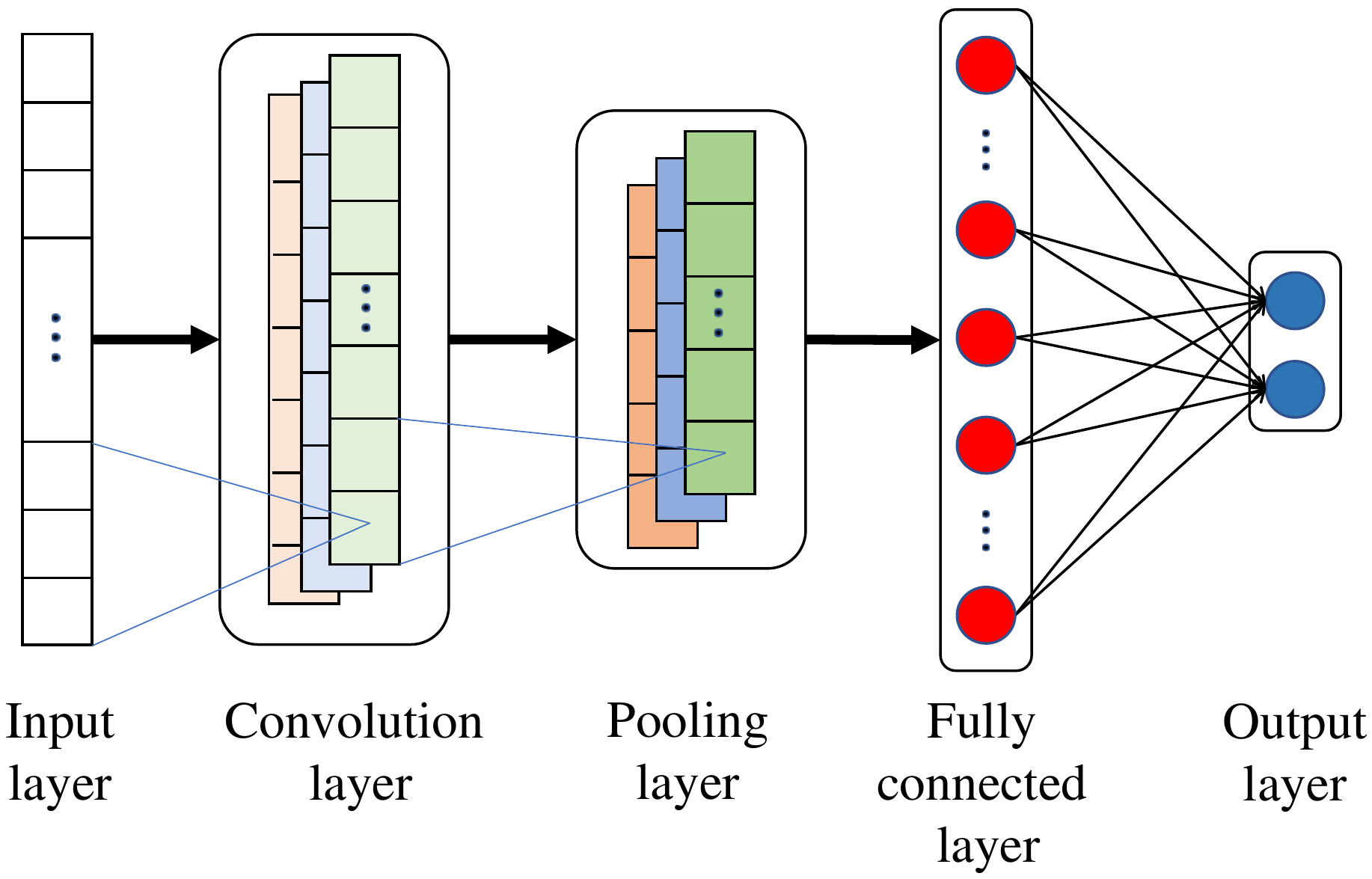}
\caption{Typical architecture of a convolutional neural network (CNN).} \label{fig:CCN_arch}
\end{figure}

\subsection{Convolutional Neural Network (CNN)}
CNN is widely used in image and natural language processing applications \cite{lecun1995convolutional} because of its capability to extract the important features and capture complex patterns in the input data. 
As shown in Fig. \ref{fig:CCN_arch}, a CNN model's architecture includes input, convolution, pooling, fully connected, and output layers. The convolution layer has a number of filters that are used to extract features from the input, and the pooling layer reduces the dimensions of the convolution layer's output. 
Convolution and pooling layers are usually followed by one or more fully connected layers that process the features extracted to be used for prediction. 


\subsection{Gated Recurrent Unit Neural Network (GRU)}
\begin{figure}[!t]
\centering
\includegraphics[scale=0.5]{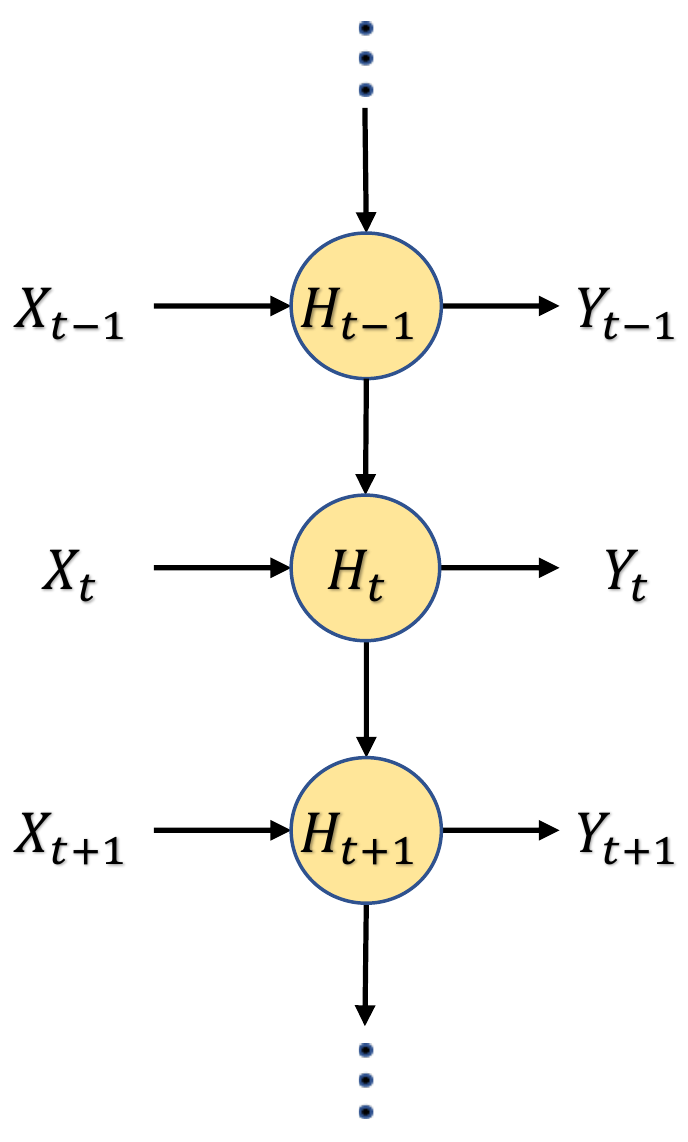}
\caption{Typical architecture of a recurrent neural network (RNN).} \label{RNN_arch}
\end{figure}

GRU is a type of RNN, which consists of hidden states and connections between the internal units to construct a directed graph as shown in Fig. \ref{RNN_arch}. 
In each time step $t$, a transition function takes the current time information $X_{t}$ and the previous hidden state ${H}_{t-1}$ to update the current hidden state ${H}_{t}$ as follows.

\begin{equation}\label{RNN_eq}{H}_{t}={F}\left({X}_{t}, {H}_{t-1}\right),\end{equation}
where ${F}$ is a nonlinear AF, e.g., Tanh. As can be observed in Eq.~\ref{RNN_eq}, ${H}_{t-1}$ can be considered as a memory for previous inputs, and thus, GRU can memorize long sequences of input patterns.
In GRU, reset and update gates are used to learn which information is important to keep and which information can be discarded. 
Therefore, GRU has the ability to capture the correlations between the inputs.
GRU is widely used in the text generation and speech recognition and synthesis applications~\cite{8985885,6638947}.

\subsection{Activation Functions (AFs)}
The AF is an important component in the machine learning models since it has a major impact on the model accuracy and convergence speed. Non-linear AFs are usually used because they enable the model to create complex mappings between the inputs and outputs.
In the following, we explain two of the AFs used in this paper~\cite{nwankpa1811activation}. 

\begin{itemize}
    \item Rectified Linear Unit (ReLU): It uses a simple max function to determine the output of a given input $x$ as follows.
    \begin{equation}
        \text{ReLU}(x) = max(0,x)
    \end{equation} 
    \item Softmax: It is commonly used in the output layer for classification problems. For a given input vector $\mathbf {z}=[z_1,\ldots , z_N]\in \mathbb {R} ^{N}$, the Softmax function is defined as follows.
\begin{equation}
    \text{Softmax} (z_i)={\frac {e^{z_i}}{\sum _{j=1}^{N}e^{z_j}}} \  \ for \ \ i = \{1, \dots, N\},
     \label{eq:1}
\end{equation}
where $N$ is the number of classes.

\end{itemize}
\section{Dataset Preparation}\label{prepare}
Due to the unavailability of a public dataset that contains both benign samples (true readings) and malicious samples (false readings) for the net metering system, we explain in this section how we prepare the dataset used in this paper. 

\subsection{Benign Readings} \label{datacorrelation}

In this paper, we use a publicly available dataset released by Ausgrid, the largest distributor of electricity on Australia's east coast \cite{ausgrid} to prepare our dataset. The Ausgrid dataset contains real power consumption and generation readings for a group of customers who are located in Sydney and regional New South Wales, and have solar panels installed on the rooftops of their homes. These readings are recorded for the period from 1-July-2010 to 30-June-2013.
Each customer has two SMs; one SM is used for measuring the power consumption and the other SM is used for measuring the generated power from the solar panels. 
The Ausgrid dataset contains information about the generation capacity that indicates the maximum amount of electricity generated from the solar panels of each customer per hour ($C_{max}$). The dataset also contains the location of each customer, the category that indicates whether an SM reading is consumption or generation, the date, and the SMs readings at half-hour granularity.  

Given the Ausgrid dataset, we apply the following operations to create our benign dataset.
\begin{itemize}
\item First, we follow the same methodology of \cite{cleanedDataset} to remove the anomalous measurements from the Ausgrid dataset and produce a clean dataset. This is because some factors such as the failure of the solar generation system can cause anomalies in the dataset.

\item Second, for each customer, we subtract the readings of the generation SM from the readings of the consumption SM to obtain net readings. These readings are equivalent to the readings that would be recorded if the two SMs were replaced by a single net meter of the net metering system because the amount of drawn/injected power from/to the utility at any time is equal to the difference between the power consumed and the power generated by the customer at this time.

\item Third, from the half-hour granularity dataset, we have created a dataset at one-hour granularity (i.e., 24 readings per day) by aggregating the readings. The reason we selected one-hour granularity is that, the lower the sampling rate, the less likely private information about the customer can be revealed \cite{jokar}. We will also show later that our detector can detect false-reading attacks with high detection rate at this reduced sampling rate.
\end{itemize}

\begin{figure}[!t]
    \centering
    \includegraphics[width=\columnwidth]{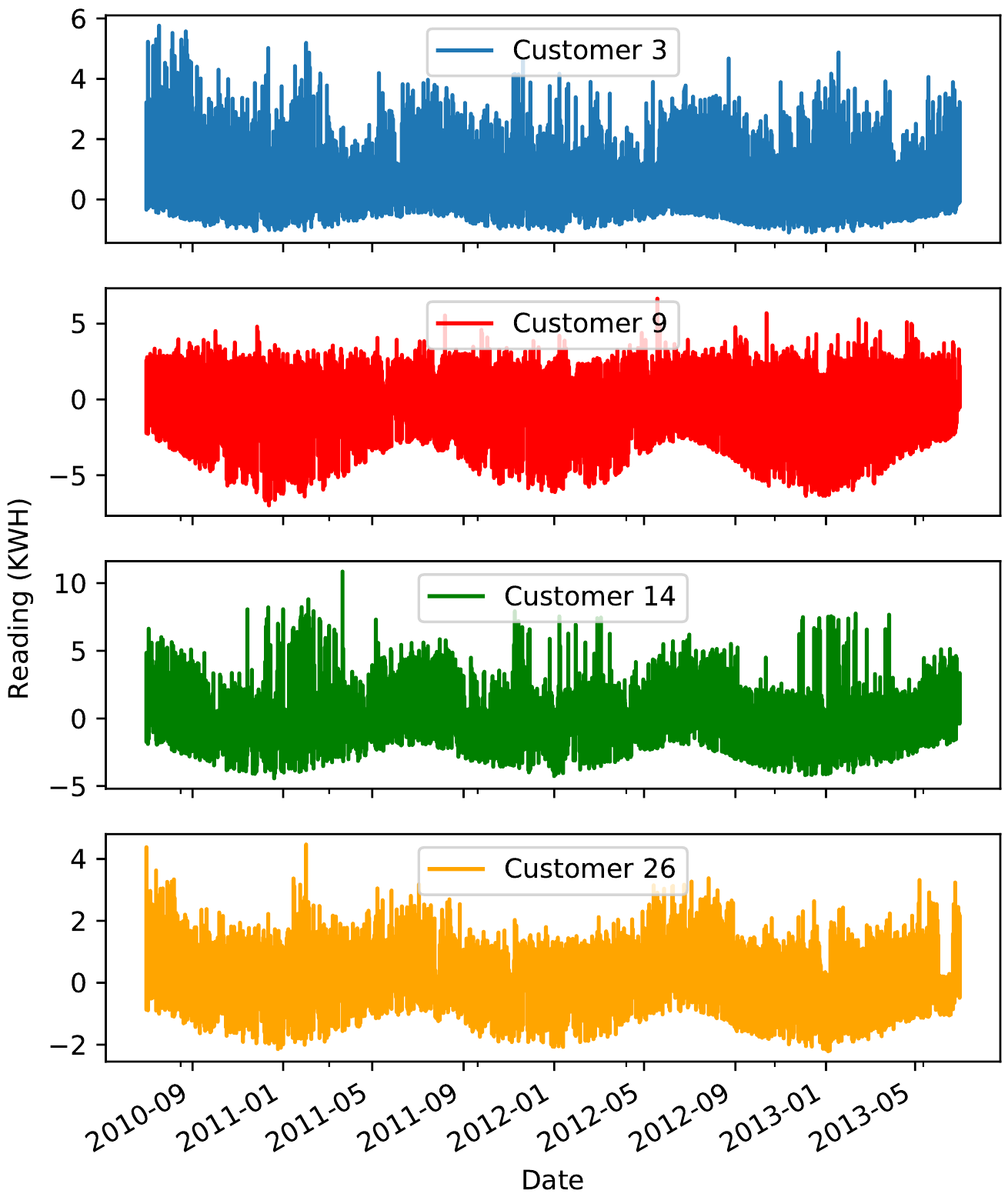}
    \caption{The net meter readings of four randomly selected customers.}
    \label{patten}
\end{figure}

Using these operations, we have created a clean dataset for 31 customers with net meter readings at one-hour granularity for 1096 days from 1-July-2010 to 30-June-2013. To get better insights from this dataset, we visualize it. Data visualization is a means that can be used to better understand the dataset by displaying the data in a visual context so that patterns and correlations within the data can be explored. 

The net meter readings of four randomly selected customers from the dataset are visualized in Fig. \ref{patten}. We can observe from the figure that the net meter readings can be positive or negative depending on the direction of the electricity flow between the customer and the utility. More importantly, we can observe that the pattern of the readings of each customer has a periodicity. For example, the shapes of the pattern of each customer over the months (from 9 to 1) are almost the same regardless of the year. We have the same observation over the months (from 1 to 5) and the months (from 5 to 9). It can also be observed that the readings are different between the days within any period.
This indicates that the readings depend on the day and the season of the year because both the consumption pattern and the amount of power generated by the solar panels depend on the season and the power consumption also depends on the day. 

\begin{figure}[!t]
    \centering
    \includegraphics[width=\columnwidth]{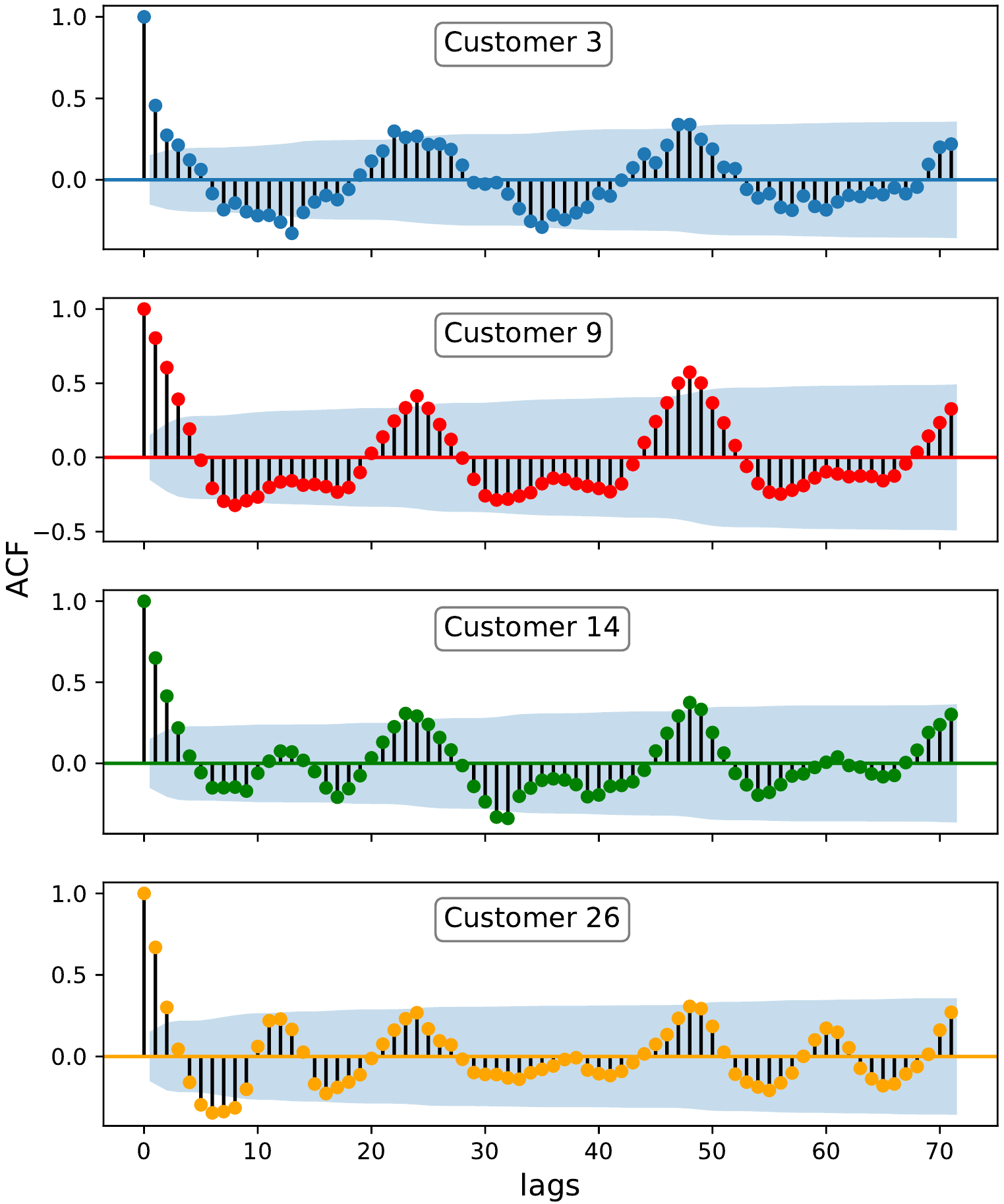}
    \caption{The ACFs of the time series data representing the net meter readings of four randomly selected customers.}
    \label{auto}
\end{figure}
\begin{table*}[!t]
\centering
\small

\caption{Proposed Attacks.}
\label{attacks}
\renewcommand{\arraystretch}{2.3}
\resizebox{\textwidth}{!}{%
\begin{tabular}{|
>{\columncolor[HTML]{C0C0C0}}c |
>{\columncolor[HTML]{C0C0C0}}c |
>{\columncolor[HTML]{C0C0C0}}c |c|c|}
\hline
\textbf{\#} &
  \multicolumn{2}{c|}{\cellcolor[HTML]{C0C0C0}\textbf{Attack}} &
  \cellcolor[HTML]{C0C0C0}\textbf{Consumption \textgreater~Generation (+ve Readings)} &
  \cellcolor[HTML]{C0C0C0}\textbf{Consumption \textless~Generation (-ve Readings)} \\ \hline\hline
\textbf{1} &
  \multicolumn{2}{c|}{\cellcolor[HTML]{C0C0C0}\textbf{Intermittent}} &
  $\left\{\begin{array}{ll}b_{t} * \text{TR}_{t},  &  t_{s} \le t \le t_{e} \\\text{TR}_{t},  &  \text {Otherwise}\end{array}\right.$ &
  $\left\{\begin{array}{l}-\max \left(p_{t} * C_{max}, \mid \text{TR}_{t}\mid\right)\!, ~t_{s} \le t \le t_{e} \\\text{TR}_{t}, \quad \text { Otherwise }\end{array}\right.$ \\ \hline
\textbf{2} &
  \cellcolor[HTML]{C0C0C0} &
  \cellcolor[HTML]{C0C0C0} &
  $\alpha* \text{TR}_{t}$ &
  $-\min (\mid \beta * \text{TR}_{t} \mid, C_{max})$ \\ \hhline{|>{\arrayrulecolor{black}}-|>{\arrayrulecolor[HTML]{C0C0C0}}-|>{\arrayrulecolor[HTML]{C0C0C0}}-|>{\arrayrulecolor{black}}--}
 
\textbf{3} &
  \cellcolor[HTML]{C0C0C0} &
  \multirow{-2}{*}{\cellcolor[HTML]{C0C0C0}\textbf{Scaling-based}} &
  $\alpha_{t} *\text{TR}_{t}$ &
  $-\min (\mid \beta_t * \text{TR}_{t} \mid, C_{max})$ \\ 
  \hhline{|>{\arrayrulecolor{black}}-|>{\arrayrulecolor[HTML]{C0C0C0}}-|>{\arrayrulecolor{black}}---}
\textbf{4} &
  \multirow{-3}{*}{\cellcolor[HTML]{C0C0C0}\textbf{Continuous}} &
  \textbf{History-based} &
  $M_{1_{t}} * \min(\text {PR}, \text{TR}_{t})$ &
  $-M_{2_{t}} * \max(\mid\text {NR}\mid, \mid\text{TR}_{t}\mid)$ \\ \hline
\end{tabular}%
}
\end{table*}

Moreover, to perceive the inner relation between the consecutive readings of the time series data representing the net meter readings of a customer, we use the autocorrelation function (ACF). The ACF gives the autocorrelation, i.e., the correlation between a time series data and itself at different time lags.
Fig. \ref{auto} shows the ACFs of the readings of the same customers selected in Fig. \ref{patten}. The shaded blue areas of Fig. \ref{auto} are the 95\% confidence intervals used to determine the significance of the autocorrelation at certain time lag. We can see that at least the autocorrelation values at time lags of 1 and 2 for all customers are located outside the blue shaded area, which indicates that there is a significant correlation between each reading and its subsequent two readings within the time series.
Furthermore, we can observe from Fig. \ref{auto} that the shape of ACF over one day is nearly similar to its shape over the consecutive days for all the customers, which indicates that the net meter readings of any customer have a daily pattern. \textit{This pattern and the time correlations between the readings can be learned by the detector to identify the false-reading attacks because deviation from the pattern can be detected as anomaly}.

\subsection{False Readings} \label{mali}
Due to the unavailability of real malicious samples for the net metering system, we propose a set of attacks to mimic the behavior of malicious customers. These attacks are given in Table \ref{attacks}. 
For attackers to achieve financial gains in the net-metering system, they should reduce their reported net readings when the power consumed is more than the power generated (i.e., the readings are positive), and increase the reported readings when the power consumed is less than the power generated (i.e., the readings are negative).
Further, the proposed attacks can be classified into \textit{intermittent} and \textit{continuous}. In \textit{intermittent} attacks, the attacker reports false readings at some time slots, and reports the true readings at other time slots aiming to confuse the detector, and in \textit{continuous} attacks, the attacker reports false readings all the time aiming to achieve high profit. 

Under the \textit{intermittent} attacks, we propose attack \#1 in which the attacker cheats during a random time interval starts at $t_s$ and ends at $t_e$, and otherwise he reports the true readings. During the cheating interval, the attacker reports a scaled-down version of the current true reading ($\text{TR}_{t}$) by a time-dependant factor $b_t$ at the time slots of positive readings, while reports the higher value between a large percentage ($p_t$) of the  maximum solar generation capacity ($C_{max}$) and the absolute value of the current true reading ($\mid\! \text{TR}_{t}\! \mid$) at the time slots of negative readings. 

Under the \textit{continuous} attacks, we propose three attacks that are either \textit{scaling-based} or \textit{history-based}. In the \textit{scaling-based} attacks, the attacker scales positive readings down and scales negative readings up without considering the values of previous readings. However, in the \textit{history-based} attack, the attacker uses the previous readings to compute the false reading. In attack \#2, the attacker cheats by always reporting a scaled-down version of $\text{TR}_{t}$ by $\alpha$ when readings are positive, while reporting a scaled-up version of $\text{TR}_{t}$ by $\beta$ when readings are negative, where $0 \le \alpha < 1$ and $\beta > 1$. Note that attack \#2 considers that the attacker's reported reading does not exceed $C_{max}$ and this is denoted by $-\min (\mid\!\!\beta * \text{TR}_{t}\!\!\mid, C_{max})$ in Table \ref{attacks}. Attack \#3 is also a \textit{scaling-based} attack, but unlike attack \#2, both the scaling down and the scaling up parameters ($\alpha$ and $\beta$) are time-dependent.

Finally, attack \#4 is \textit{history-based} in which the attacker cheats by reporting the minimum value between $\text{TR}_{t}$ and last reported positive reading ($\text {PR}$) when readings are positive, and reporting the maximum value between $\text{TR}_{t}$ and last reported negative reading ($\text {NR}$) when readings are negative. Note that the factors $M_{1_{t}}$ and $M_{2_{t}}$ in attack \#4 are not scaling factors but they act as masks to avoid reporting the same exact reading in multiple time slots to confuse the detector; where the value of $M_{1_{t}}$ is a little bit less than one, while  the value of $M_{2_{t}}$ is a little bit more than one.

\begin{figure*}[!t]
    \centering
    \includegraphics[scale=0.5]{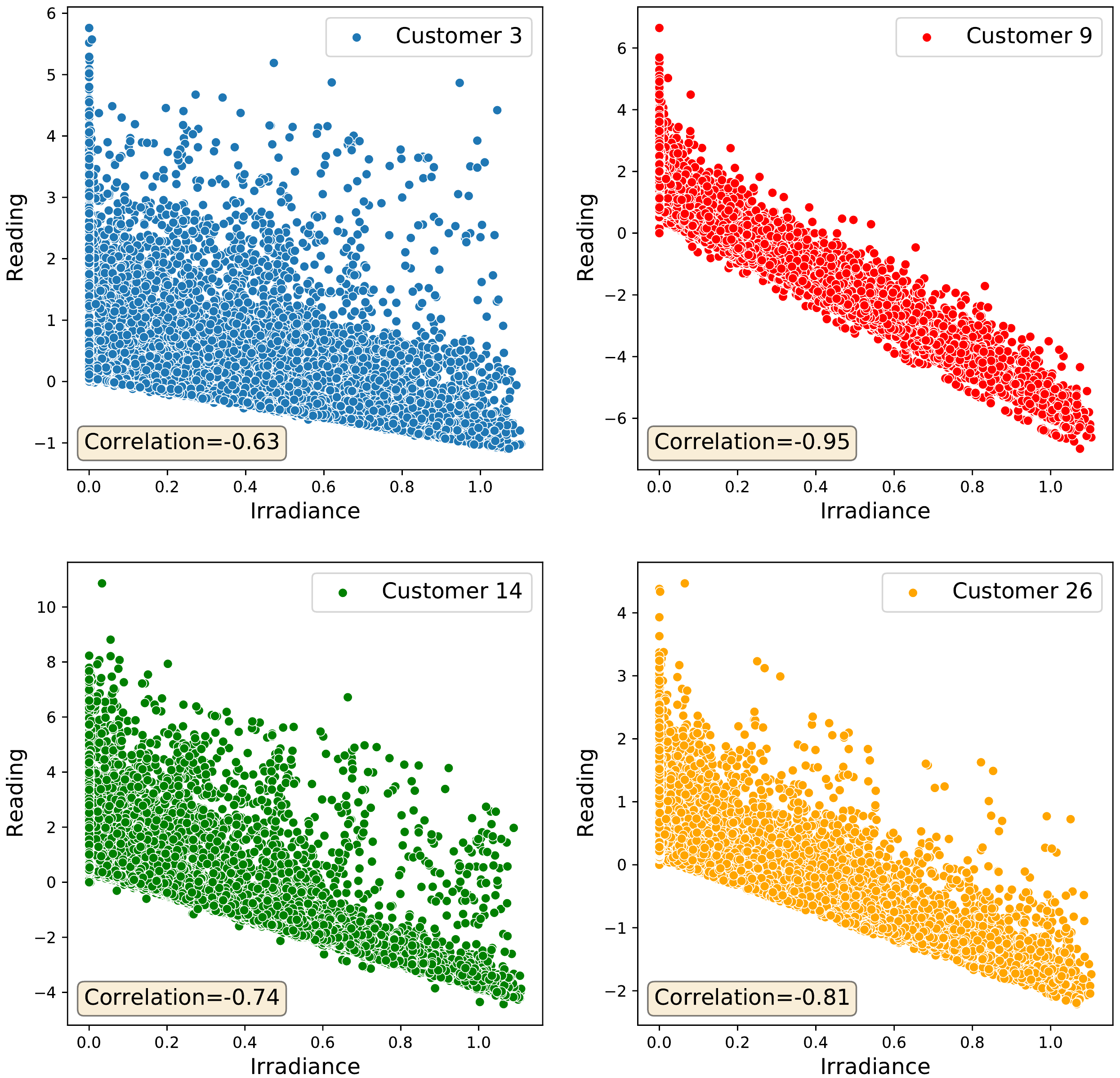}
    \caption{The correlation between the net meter readings and the solar irradiance for four randomly selected customers.}
    \label{corro}
\end{figure*}

\subsection{Relevant Data from Trustworthy Sources} \label{other}
According to \cite{generationEquations}, the amount of power generated from a solar panel depends on both the solar irradiance and the temperature. Moreover, $C_{max}$ of each customer can be calculated by the utility based on the number of solar panels and their characteristics recorded in the contract between the customer and the utility. Therefore, in addition to using the readings reported by the net meter to detect the false-reading attacks, relevant data from trustworthy sources, including the solar irradiance, temperature, and $C_{max}$, can be used by the detector because they give indication about the power generation pattern that can be used to verify the net meter readings. 
The rationale here is that although an attacker can compromise his net meter to report false readings, he cannot manipulate the solar irradiance, temperature, and $C_{max}$ because they are beyond his control. Thus, \textit{the true data from the trustworthy sources can help the detector to identify the false-reading attacks}.

The Ausgrid dataset contains $C_{max}$ of each customer but it does not contain information about the solar irradiance and temperature. However, they can be obtained from SOLCAST \cite{solcast} using the locations of customers given in Ausgrid. SOLCAST is a website that can provide the weather information including the solar irradiance and temperature of any location in the world at any date given the longitude and the latitude of this location. Thus, given the customers' locations, we have found the longitudes and latitudes of these locations to obtain the solar irradiance and temperature at these locations during the period from 1-July-2010 to 30-June-2013 via SOLCAST.

Given the time series data representing the net meter readings of a certain customer and the time series data representing the corresponding values of the solar irradiance, the correlation between the two time series data is visualized in Fig. \ref{corro} for the same customers selected in Figs. \ref{patten} and \ref{auto}. The scatter plots of Fig. \ref{corro} indicate a negative correlation between the net meter readings and the solar irradiance. This is because the higher the irradiance, the higher the generated power by the solar panels and thus the lower the reported net reading. Furthermore, the correlation values given in each subplot of Fig. \ref{corro} indicate the significance of this correlation for all customers. Similarly, we have used a similar approach but with using the temperature and found also a negative correlation between the net meter readings and the temperature. Therefore, \textit{if the readings reported by the net meter are true, there should be correlations between the readings and the values of the irradiance and temperature}, otherwise, the detector can decide that the reported readings are false.

\subsection{Data Preprocessing}
Given the dataset of benign net meter readings of 31 customers for 1096 days, the readings of each day are treated as a benign sample for a total of 33,976 (31*1096) benign samples. Then, the four proposed attacks are used to create four malicious samples from each benign sample. After that, our dataset of benign and malicious samples is extended by including more data. Specifically, each sample in the extended dataset consists of 75 values representing 24 fine-grained net meter readings,  the corresponding 24 fine-grained irradiance values, the corresponding 24 fine-grained temperature values, $C_{max}$, the day, and the season.
The dataset that is further divided into training and test sets with a ratio of 2:1, respectively. After that, the training and test sets are normalized to bring all the features' values to a common scale to guarantee the fair contributions of all the features towards the classification of the detector. This step is also useful to help the detectors that use the Gradient Descent  optimization to converge faster \cite{normalization}. Finally, given that the number of malicious samples is four times the number of benign samples, the adaptive synthetic (ADASYN) sampling approach \cite{adasyn} is used to balance the training set by over-sampling the minority class of benign samples to avoid biasing the trained model towards the majority class of malicious samples.

\begin{figure*}[!t]
    \centering
\includegraphics[scale=0.59]{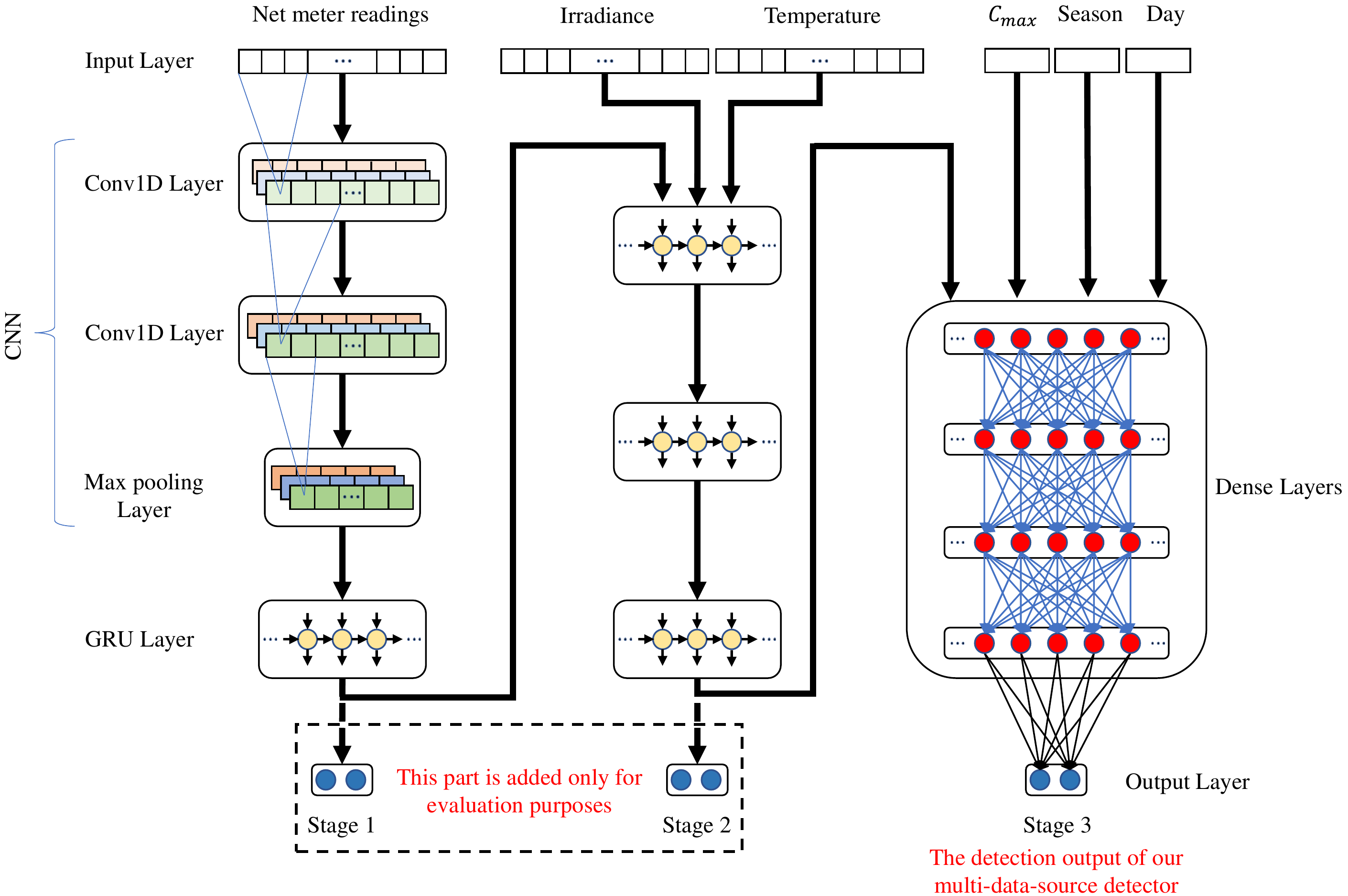}
    \caption{The architecture of our false-reading attacks detector.}
    \label{proposedArchitecture}
\end{figure*}

\section{Proposed Detector for False-Reading Attacks} \label{chap:design} 

In this section, we first discuss the rationale behind the design of the proposed detector that detects the false-reading attacks. Then, we describe the architecture of the proposed detector in detail.

\subsection{The Rationale Behind the Detector Design} 
\textbf{Machine learning}. Among the existing detectors of false-reading attacks in the AMI network, machine learning-based detectors outperform the other detectors that are based on state estimation and game theory \cite{nabil2018}.\\
\textbf{Deep learning}. The machine learning-based detectors can use either shallow classifiers such as DT, RF and SVM or deep learning architectures such as the CNN and the RNN \cite{kenz2016}. However, the results of recent studies have indicated that the deep learning architectures can accurately detect false-reading attacks in the AMI network better than the shallow classifiers \cite{kenz2016,course, espain,ismail2020}.\\
\textbf{General detector}. The detector can be either customer-specific, where a customized detector is trained for each customer or a general detector that can be used for all customers. Unlike general detectors, customer-specific detectors require collecting historical metering readings for each customer to train them, and thus, they cannot be used to detect false-reading attacks launched by new customers until enough readings are collected. Besides, customer-specific detectors are vulnerable to contamination attacks, where a new customer reports false readings from the beginning. If the detector is trained on this data, the customer can continue reporting false data without being detected  \cite{nabil2018}.\\
\textbf{Data correlation}.  Based on the data analysis provided in Section \ref{datacorrelation}, there are time correlations between the consecutive readings within the benign samples of any customer. Thus, it is important for the detector to be able to learn these correlations so that it can identify false-reading attacks if these correlations are not found in the tested sample.\\
\textbf{Multi-source data}. Based on the data analysis provided in Section \ref{other}, there are correlations between the time-series data representing the true net meter readings and the time-series data obtained from trustworthy sources such as the solar irradiance and temperature that are always true because they are beyond the control of customers. 
Thus, it is good for the detector to learn these correlations so that it can identify false-reading attacks if these correlations are not found in the tested sample of net meter readings and the corresponding values of the solar irradiance and temperature.
Moreover, we have shown in Section \ref{prepare} that data such as the day, the season, and $C_{max}$ are also important to be considered because the net meter readings depend on the day and the season, and $C_{max}$ can help the detector to perceive the limits of the reported readings.

Based on the above discussion, our detector shall have the following characteristics. It shall be a general deep learning-based detector trained on the data collected from all customers to be used for any customer in the system including the new customers. Moreover, the deep learning architecture of the detector needs to capture the correlation within the net meter readings.
Finally, instead of designing a single-data-source detector that detects false-reading attacks based solely on the readings reported by the net meter, our detector shall be a multi-data-source detector that is trained on relevant data from various trustworthy sources in addition to the net meter readings to enhance the  detection performance.

\subsection{The Architecture of Our Detector}
Our detector is designed to have six different types of data (i.e., data from six different sources) as shown in Fig. \ref{proposedArchitecture}. The first one is the fine-grained net meter readings of one day. The second and third input data are the fine-grained irradiance and temperature values in the same day, respectively. The rest of the input data are the values of  $C_{max}$, the day , and the season.
Moreover, we have designed our detector in three stages as shown in Fig. \ref{proposedArchitecture} to enhance the detection performance of our detector by considering more input data at each stage. These stages are described as follows.
\begin{itemize}

\item Stage 1 considers only one type of input data, which is the net meter readings, and it has a hybrid CNN \& GRU architecture. Since the time-series data can be considered as 1-dimensional (1-D) data, the 1-D CNN architecture can be used due to its capability to extract the features from the 1-D data and hence leading to a high detection performance. Moreover, since the time-series data representing the net meter readings has correlations between the consecutive readings, the GRU architecture can be used due to its capability to capture the time correlation in the data.  Thus, we have chosen a hybrid CNN \& GRU architecture for Stage 1 so that the CNN layers can extract the features from the input net meter readings while the GRU layers can capture the correlation between the extracted features.

\item Stage 2 considers three input data by feeding it with the output of Stage 1, the irradiance, and the temperature. Stage 2 has a set of GRU layers to help the detector to capture the correlation between the net meter readings and the corresponding values of the irradiance and the temperature. 

\item Stage 3 takes all the six input data into consideration by feeding it with the output of Stage 2, $C_{max}$, the day and the season. Stage 3 has a set of dense layers to help the detector to take complex decisions regarding the input samples by considering the effect of important features like $C_{max}$, the day and the season on the output of Stage 2 that represents the net meter reading, the irradiance, and the temperature.
\end{itemize}

In order to assess the impact of considering more input data on the detection performance, in the evaluation we get an output from each stage as shown in Fig. \ref{proposedArchitecture}, but when the detector is used by the utility, the output should be obtained from Stage 3 only.

\section{Performance Evaluation} \label{performanceEvaluation}
In this section, we first discuss our experimental environment and the performance evaluation metrics. Then, we present two experiments we have conducted to evaluate the performance of our detector. In the first experiment, we investigate various deep learning architectures to detect false-reading attacks based solely on the readings reported by the net meters. In the second experiment, we investigate the enhancement in the detection performance of the detector due to considering relevant data from trustworthy sources besides the reported readings. 

\subsection{Experimental Setup and Evaluation Metrics} \label{met}
In this subsection, we describe the details of our experimental environment in terms of software and hardware, and also the evaluation metrics used to assess the performance of our detector.

Various Python 3 libraries are used in our work as follows. Specifically, Pandas and Numpy are used in data preparation, while Matplotlib \cite{matplotlib}, Statsmodels \cite{statsmodels}, and Seaborn \cite{seaborn} are used in data visualization. To train the detectors and optimize the hyper-parameters, Keras Functional API \cite{keras} and Hyperopt \cite{hyperopt} are used, respectively. Finally, Sklearn \cite{scikit-learn} is used for evaluating the performance of the detector. All the experiments are run on the high-performance cluster of the Tennessee Technological University using two NVIDIA Tesla K80 GPUs.

The following metrics are considered to evaluate the performance of our detector.
\begin{itemize}
    \item \textbf{Accuracy ($ACC$)}. It measures the percentage of the correctly classified samples to the total number of tested samples, and it is calculated as follows.
   $$ACC (\%)=\frac{TP+TN}{TP+TN+FP+FN}\times100,$$
   where, $TP$ is the number of true positive samples (i.e, the correctly classified malicious samples), $TN$ is the number of true negative samples (i.e, the correctly classified benign samples), $FP$ is the number of false positive samples (i.e, the misclassified benign samples) and $FN$ is the number of false negative samples (i.e, the misclassified malicious samples).
   
   \item \textbf{Precision ($PR$)}. It measures the percentage of the correctly classified positive samples to the total number of samples classified as positive and it is calculated as follows.
   $$PR (\%)=\frac{TP}{TP+FP}\times100$$
   
   \item \textbf{Detection rate ($DR$)}. It measures the percentage of the correctly classified positive samples to the total number of real positive samples and it is calculated as follows.
   $$DR (\%)=\frac{TP}{TP+FN}\times100$$
   
   \item \textbf{False Alarm ($FA$)}. It measures the percentage of misclassified negative samples to the total number of real negative samples and it is calculated as follows.
   $$FA (\%)=\frac{FP}{FP+TN}\times100$$
   
   \item \textbf{Highest difference ($HD$)}. It measures the difference between $DR$ and $FA$.
   $$HD (\%)=DR(\%) - FA(\%)$$
   
   \item \textbf{F1-score ($F1$)}. It is the harmonic mean between $PR$ and $DR$ and it is calculated as follows.
  {$$F1 (\%)=\frac{2*PR*DR}{PR+DR}\times100,$$}
   
    \item \textbf{Receiver operating characteristic (ROC) curve}. It is a graphical representation of the relation between $TP$ rate and $FP$ rate at different decision thresholds. The higher the area under the ROC curve (AUC-ROC), the higher the detector performance.
    
    \item \textbf{Precision-recall (P-R) curve}.  It is a graphical representation of the relation between $PR$ and recall at different decision thresholds. The higher the area under the P-R curve (AUC-P-R), the higher the detector performance.
\end{itemize}

\subsection{Results of Experiment 1} \label{inves}
In this subsection, we investigate four possible deep learning architectures, namely, MLP, GRU, CNN, and hybrid CNN \& GRU, to determine the one that provides the best performance by considering only the readings reported by the customers' net meters.
The MLP is firstly investigated since it is the simplest deep learning architecture, and then iw will be used as a baseline for assessing the performance of the other deep learning architectures. 
To do a fair investigation, the dataset prepared in Section \ref{prepare} is used to train four different detectors with the aforementioned architectures, and Hyperopt is used to optimize the hyper-parameters including the number of layers, the number of units per layer, and the AF used in each layer. The optimal hyper-parameters of these detectors are given in Tables \ref{hyper1}-\ref{hyper4}.

\textbf{Results and Discussion}.
Table \ref{compare} gives a comparison between the performance of the four detectors in terms of $ACC$, $PR$, $DR$, $FA$, $HD$, and $F1$. First, we can see that while the MLP-based detector has the lowest computational complexity among all detectors, it achieves the lowest performance. Second, it can be observed that the GRU-based detector provides a better performance compared to the MLP-based detector, which makes sense because the GRU layers are capable of capturing the correlation between the inputs and it has been proved in Section \ref{datacorrelation} that there is a temporal correlation between the consecutive net meter readings. Third, it can be observed that the CNN-based detector provides a better performance compared to the MLP-based detector, which makes sense because the CNN layers provide the detector with a better feature extraction capability. Overall, the hybrid CNN \& GRU-based detector achieves the best performance among all detectors because the CNN layers can extract the features from the input readings while the GRU layers can capture the correlation between the extracted features. Thus, the hyprid  CNN \& GRU architecture is chosen to design our multi-data-source detector.

\begin{table}[!t]
\caption{The optimal hyper-parameters of the MLP-based detector.}
\label{hyper1}
\centering
\renewcommand{\arraystretch}{1.3}
\resizebox{\columnwidth}{!}{%
\tiny
\begin{tabular}{|c|c|c|c|}
\hline
\rowcolor[HTML]{C0C0C0} 
\cellcolor[HTML]{C0C0C0}                               & \multicolumn{3}{c|}{\cellcolor[HTML]{C0C0C0}\textbf{Hyper-parameters~~~}} \\ 
\hhline{|>{\arrayrulecolor[HTML]{C0C0C0}}-|>{\arrayrulecolor{black}}---}
\rowcolor[HTML]{C0C0C0} 
\multirow{-2}{*}{\cellcolor[HTML]{C0C0C0}\textbf{Architecture}} & \textbf{Layer}           & \textbf{Number of units}          & \textbf{AF}               \\ \hline\hline
\cellcolor[HTML]{C0C0C0}                               & Input           & 24                       & Linear           \\ \cline{2-4} 
\cellcolor[HTML]{C0C0C0}                               & Dense           & 128                      & Linear           \\ \cline{2-4} 
\cellcolor[HTML]{C0C0C0}                               & Dense           & 128                      & Sigmoid          \\ \cline{2-4} 
\cellcolor[HTML]{C0C0C0}                               & Dense           & 128                      & Sigmoid          \\ \cline{2-4} 
\cellcolor[HTML]{C0C0C0}                               & Dense           & 256                      & Sigmoid          \\ \cline{2-4} 
\cellcolor[HTML]{C0C0C0}                               & Dense           & 256                      & Relu             \\ \cline{2-4} 
\cellcolor[HTML]{C0C0C0}                               & Dense           & 256                      & Elu              \\ \cline{2-4} 
\multirow{-8}{*}{\cellcolor[HTML]{C0C0C0}\textbf{MLP}}          & Output          & 2                        & Softmax          \\ \hline
\end{tabular}%
}
\end{table}

 \begin{table}[!t]
\caption{The optimal hyper-parameters of the GRU-based detector.}
\label{hyper3}
\centering
\renewcommand{\arraystretch}{1.3}
\resizebox{\columnwidth}{!}{%
\tiny
\begin{tabular}{|c|c|c|c|}
\hline
\rowcolor[HTML]{C0C0C0} 
\cellcolor[HTML]{C0C0C0}                               & \multicolumn{3}{c|}{\cellcolor[HTML]{C0C0C0}\textbf{Hyper-parameters~~~}} \\ \hhline{|>{\arrayrulecolor[HTML]{C0C0C0}}-|>{\arrayrulecolor{black}}---} 
\rowcolor[HTML]{C0C0C0} 
\multirow{-2}{*}{\cellcolor[HTML]{C0C0C0}\textbf{Architecture}} & \textbf{Layer}           & \textbf{Number of units}          & \textbf{AF}               \\ \hline\hline
\cellcolor[HTML]{C0C0C0}                               & Input           & 24                       & Linear           \\ \cline{2-4} 
\cellcolor[HTML]{C0C0C0}                               & GRU             & 64                       & Sigmoid          \\ \cline{2-4} 
\cellcolor[HTML]{C0C0C0}                               & GRU             & 128                      & Relu             \\ \cline{2-4} 
\multirow{-4}{*}{\cellcolor[HTML]{C0C0C0}\textbf{GRU}}          & Output          & 2                        & Softmax          \\ \hline
\end{tabular}%
}
\end{table}

\begin{table}[!t]

\caption{The optimal hyper-parameters of the CNN-based detector.}
\label{hyper2}
\centering
\renewcommand{\arraystretch}{1.3}
\resizebox{\columnwidth}{!}{%
\tiny
\begin{tabular}{|c|c|c|c|}
\hline
\rowcolor[HTML]{C0C0C0} 
\cellcolor[HTML]{C0C0C0}                               & \multicolumn{3}{c|}{\cellcolor[HTML]{C0C0C0}\textbf{Hyper-parameters}} \\ \hhline{|>{\arrayrulecolor[HTML]{C0C0C0}}-|>{\arrayrulecolor{black}}---} 
\rowcolor[HTML]{C0C0C0} 
\multirow{-2}{*}{\cellcolor[HTML]{C0C0C0}\textbf{Architecture}} & \textbf{Layer}           & \textbf{Number of units}          & \textbf{AF}               \\ \hline\hline
\cellcolor[HTML]{C0C0C0} & Input  & 24  & Linear  \\ \cline{2-4} 
\cellcolor[HTML]{C0C0C0} & Conv1D & 128 & Relu    \\ \cline{2-4} 
\cellcolor[HTML]{C0C0C0} & Conv1D & 64  & Tanh    \\ \cline{2-4} 
\cellcolor[HTML]{C0C0C0} & Dense  & 256 & Sigmoid \\ \cline{2-4} 
\cellcolor[HTML]{C0C0C0} & Dense  & 128 & Elu     \\ \cline{2-4} 
\cellcolor[HTML]{C0C0C0} & Dense  & 128 & Tanh    \\ \cline{2-4} 
\cellcolor[HTML]{C0C0C0} & Dense  & 256 & Sigmoid \\ \cline{2-4} 
\cellcolor[HTML]{C0C0C0} & Dense  & 512 & Relu    \\ \cline{2-4} 
\cellcolor[HTML]{C0C0C0} & Dense  & 128 & Tanh    \\ \cline{2-4} 
\multirow{-10}{*}{\cellcolor[HTML]{C0C0C0}\textbf{CNN}}         & Output              & 2                             & Softmax          \\ \hline
\end{tabular}%
}
\end{table}


\begin{table}[!t]
\caption{The optimal hyper-parameters of the hybrid CNN \& GRU-based detector.}
\label{hyper4}
\centering
\renewcommand{\arraystretch}{1.3}
\resizebox{\columnwidth}{!}{%
\tiny
\begin{tabular}{|c|c|c|c|}
\hline
\rowcolor[HTML]{C0C0C0} 
\cellcolor[HTML]{C0C0C0}                               & \multicolumn{3}{c|}{\cellcolor[HTML]{C0C0C0}\textbf{Hyper-parameters}} \\ \hhline{|>{\arrayrulecolor[HTML]{C0C0C0}}-|>{\arrayrulecolor{black}}---}
\rowcolor[HTML]{C0C0C0} 
\multirow{-2}{*}{\cellcolor[HTML]{C0C0C0}\textbf{Architecture}} & \textbf{Layer}           & \textbf{Number of units}          & \textbf{AF}               \\ \hline\hline
\cellcolor[HTML]{C0C0C0} & Input  & 24 & Linear \\ \cline{2-4} 
\cellcolor[HTML]{C0C0C0} & Conv1D & 64 & Relu   \\ \cline{2-4} 
\cellcolor[HTML]{C0C0C0} & Conv1D & 32 & Relu   \\ \cline{2-4} 
\cellcolor[HTML]{C0C0C0} & GRU    & 32 & Relu   \\ \cline{2-4} 
\multirow{-5}{*}{\cellcolor[HTML]{C0C0C0}\textbf{CNN \& GRU}}   & Output              & 2                             & Softmax          \\ \hline
\end{tabular}%
}
\end{table}


\begin{table}[!t]
\caption{Comparison between the performance of the different detectors.}
\label{compare}
\centering
\renewcommand{\arraystretch}{1.3}
\resizebox{\columnwidth}{!}{%
\begin{tabular}{|
>{\columncolor[HTML]{C0C0C0}}c |c|c|c|c|c|c|}
\hline
\cellcolor[HTML]{C0C0C0} & \multicolumn{6}{c|}{\cellcolor[HTML]{C0C0C0}\textbf{Metrics}} \\ \hhline{|>{\arrayrulecolor[HTML]{C0C0C0}}-|>{\arrayrulecolor{black}}------} 
\multirow{-2}{*}{\cellcolor[HTML]{C0C0C0}\textbf{Architecture}} & \cellcolor[HTML]{C0C0C0}\textbf{ACC} & \cellcolor[HTML]{C0C0C0}\textbf{PR} &\cellcolor[HTML]{C0C0C0} \textbf{DR} & \cellcolor[HTML]{C0C0C0}\textbf{FA} &\cellcolor[HTML]{C0C0C0} \textbf{HD} & \cellcolor[HTML]{C0C0C0}\textbf{F1} \\ \hline\hline
\textbf{MLP}             & 94.53    & 98.35    & 94.35    & 6.36    & 87.99    & 96.3    \\ \hline
\textbf{GRU}             & 95.9     & 98.5     & 95.6     & 5.33    & 90.27    & 97.02   \\ \hline
\textbf{CNN}             & 94.92    & 99.01    & 94.57    & 3.89    & 90.68    & 96.57   \\ \hline
\textbf{CNN \& GRU}         & 95.94    & 99.02    & 95.74    & 3.79    & 91.83    & 97.35   \\ \hline
\end{tabular}%
}
\end{table}

\begin{table}[!t]
\caption{The optimal hyper-parameters of the stages of our detector.}
\label{hyper5}
\centering
\renewcommand{\arraystretch}{1.3}
\resizebox{\columnwidth}{!}{%
\tiny
\begin{tabular}{|c|c|c|c|}
\hline
\rowcolor[HTML]{C0C0C0} 
\cellcolor[HTML]{C0C0C0}                                 & \multicolumn{3}{c|}{\cellcolor[HTML]{C0C0C0}\textbf{Hyper-parameters}} \\ \hhline{|>{\arrayrulecolor[HTML]{C0C0C0}}-|>{\arrayrulecolor{black}}---}
\rowcolor[HTML]{C0C0C0} 
\multirow{-2}{*}{\cellcolor[HTML]{C0C0C0}\textbf{Stage}} & \textbf{Layer}      & \textbf{Number of units}      & \textbf{AF}      \\ \hline\hline
\cellcolor[HTML]{C0C0C0}                                   & Input  & 24  & Linear  \\ \cline{2-4} 
\cellcolor[HTML]{C0C0C0}                                   & Conv1D & 64  & Relu    \\ \cline{2-4} 
\cellcolor[HTML]{C0C0C0}                                   & Conv1D & 64  & Relu    \\ \cline{2-4} 
\cellcolor[HTML]{C0C0C0}                                   & GRU    & 64  & Sigmoid \\ \cline{2-4} 
\multirow{-5}{*}{\cellcolor[HTML]{C0C0C0}\textbf{Stage 1}} & Output & 2   & Softmax \\ \hline\hline
\cellcolor[HTML]{C0C0C0}                                   & Input  & 48  & Linear  \\ \cline{2-4} 
\cellcolor[HTML]{C0C0C0}                                   & GRU    & 128 & Tanh    \\ \cline{2-4} 
\cellcolor[HTML]{C0C0C0}                                   & GRU    & 64  & Tanh    \\ \cline{2-4} 
\cellcolor[HTML]{C0C0C0}                                   & GRU    & 128 & Tanh    \\ \cline{2-4} 
\multirow{-5}{*}{\cellcolor[HTML]{C0C0C0}\textbf{Stage 2}} & Output & 2   & Softmax \\ \hline\hline
\cellcolor[HTML]{C0C0C0}                                   & Input  & 3   & Linear  \\ \cline{2-4} 
\cellcolor[HTML]{C0C0C0}                                   & Dense  & 128 & Relu    \\ \cline{2-4} 
\cellcolor[HTML]{C0C0C0}                                   & Dense  & 128 & Relu    \\ \cline{2-4} 
\cellcolor[HTML]{C0C0C0}                                   & Dense  & 128 & Relu    \\ \cline{2-4} 
\cellcolor[HTML]{C0C0C0}                                   & Dense  & 64  & Relu    \\ \cline{2-4} 
\multirow{-6}{*}{\cellcolor[HTML]{C0C0C0}\textbf{Stage 3}} & Output & 2   & Softmax \\ \hline
\end{tabular}%
}
\vspace{.5ex}

Note: The input to Stage 2 is (48 for solar irradiance and temperature in addition to the output of Stage 1), and the input to Stage 3 is (3 for day, season, and $C_{max}$ in addition to the output of Stage 2).
\end{table}

\subsection{Results of Experiment 2}
In this subsection, we provide a comparison between the outputs of Stages 1, 2, and 3 of our detector shown in Fig. \ref{proposedArchitecture} to evaluate the benefit of considering multiple data sources.
To train our detector, the dataset prepared in Section \ref{prepare} is used as follows. Stage 1 is trained using only the net meter readings, While Stage 2 is trained using the readings besides the solar irradiance and temperature. Finally, Stage 3 is trained using all data included in our dataset.
In addition, Hyperopt is used to optimize the hyper-parameters of the different stages of our detector, and the optimal hyper-parameters are given in Table \ref{hyper5}.


\begin{table}[!t]
\caption{Comparison between the performance of the three stages of our detector.}
\label{compare2}
\centering
\renewcommand{\arraystretch}{1.3}
\resizebox{\columnwidth}{!}{%
\begin{tabular}{|c|c|c|c|c|c|c|}
\hline
\rowcolor[HTML]{C0C0C0} 
\cellcolor[HTML]{C0C0C0}                 & \multicolumn{6}{c|}{\cellcolor[HTML]{C0C0C0}\textbf{Metrics}} \\ \hhline{|>{\arrayrulecolor[HTML]{C0C0C0}}-|>{\arrayrulecolor{black}}------}  
\rowcolor[HTML]{C0C0C0} 
\multirow{-2}{*}{\cellcolor[HTML]{C0C0C0}\textbf{Stage}} & \textbf{ACC} & \textbf{PR} & \textbf{DR} & \textbf{FA} & \textbf{HD} & \textbf{F1} \\ \hline\hline
\cellcolor[HTML]{C0C0C0}\textbf{Stage 1} & 95.94    & 99.02    & 95.74    & 3.79    & 91.83    & 97.35   \\ \hline
\cellcolor[HTML]{C0C0C0}\textbf{Stage 2} & 96.97    & 99.14    & 97.06    & 3.35    & 93.71    & 98.09   \\ \hline
\cellcolor[HTML]{C0C0C0}\textbf{Stage 3} & 98.29    & 99.26    & 98.59    & 2.92    & 95.66    & 98.93   \\ \hline
\end{tabular}%
}
\end{table}

\begin{figure}[!t]
    \centering
    \includegraphics[width=\columnwidth]{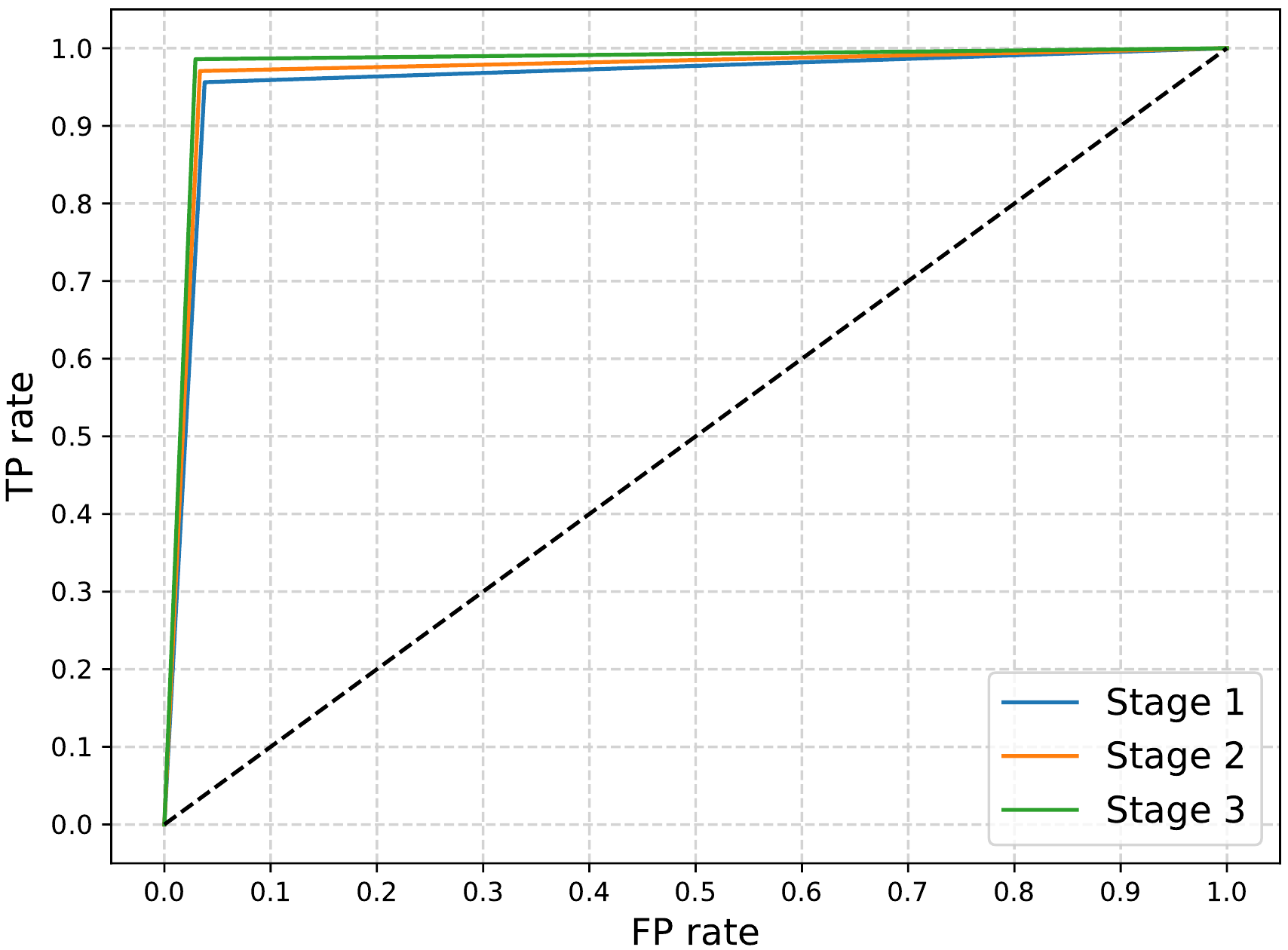}
    \caption{Comparison between the ROC curves of Stages 1, 2, and 3.}
    \label{roccurve}
\end{figure}

\begin{figure}[!t]
    \centering
    \includegraphics[width=\columnwidth]{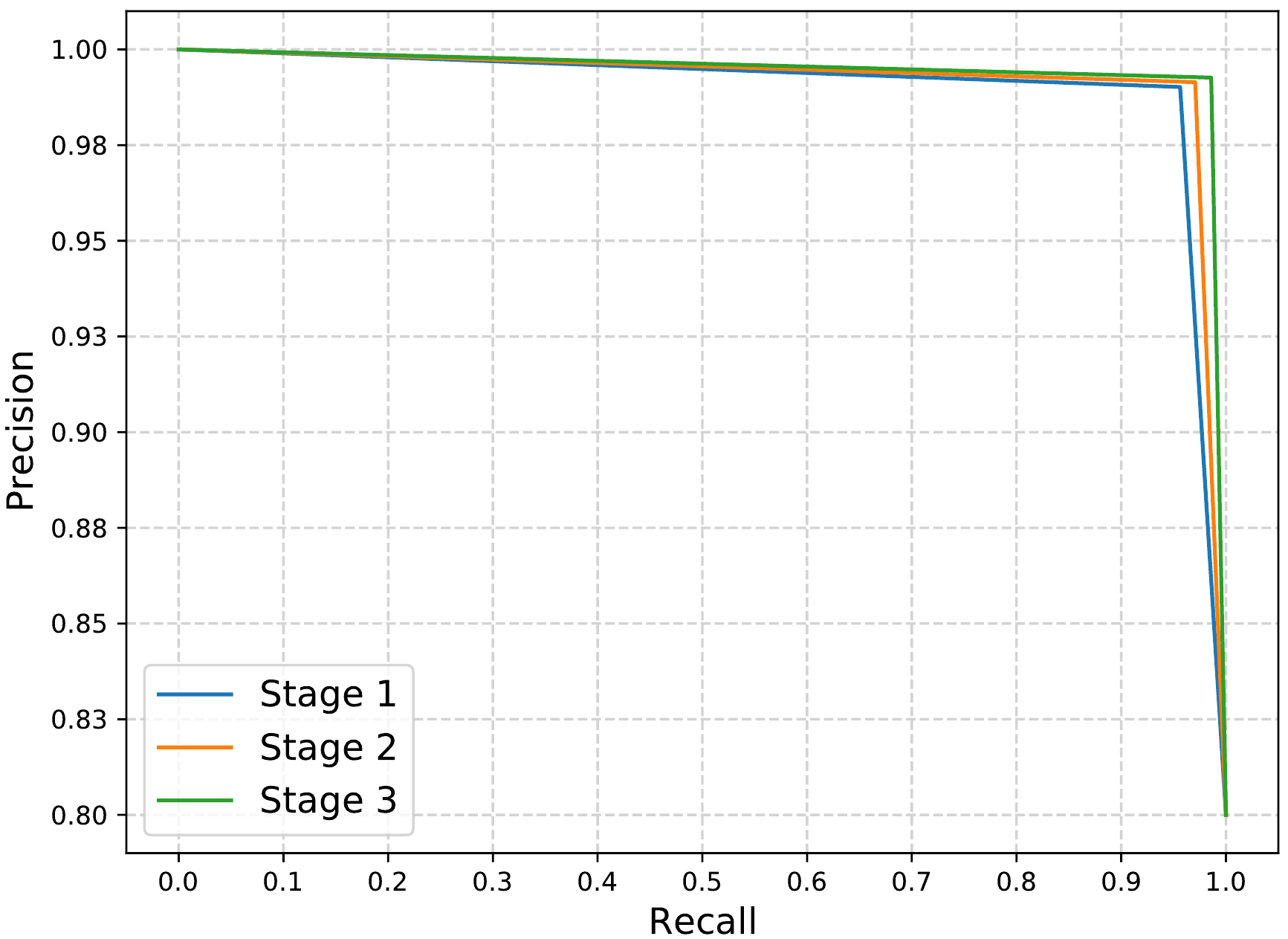}
    \caption{Comparison between the P-R curves of Stages 1, 2, and 3.}
    \label{prcurve}
\end{figure}

\textbf{Results and Discussion}.
Table \ref{compare2} gives a comparison between the performance of the three stages in terms of $ACC$, $PR$, $DR$, $FA$, $HD$ and $F1$. In addition, Figs. \ref{roccurve} and \ref{prcurve} visualize the difference in the performance between the three stages using ROC curves and P-R curves, respectively. It can be clearly concluded from the results given in Table \ref{compare2} that Stage 3 achieves the best performance among all stages. This can also be observed in Figs. \ref{roccurve} and \ref{prcurve} because Stage 3 has the biggest AUC-ROC and AUC-P-R, respectively among all stages. 

The superiority of Stage 2 over Stage 1 is due to the capability of Stage 2 to successfully capture the correlations between the net meter readings and the corresponding values of the solar irradiance and temperature. We have discussed in Section \ref{other} that these correlations are significant when the reported readings are true. Thus, if a customer maliciously manipulates his readings, this can affect the correlations between the net meter readings and the corresponding values of the irradiance and temperature. For a malicious customer to fully preserve these correlations, he has to manipulate the solar irradiance and temperature values the same way he manipulates the net meter readings. However, this is impossible because the irradiance and temperature values are not reported by him. This means that Stage 2 has additional features that help in differentiating between the benign and malicious samples by checking the correlations between the net meter readings and the corresponding values of the irradiance and temperature, which results in higher $DR$, lower $FA$, and higher $HD$.

The superiority of Stage 3 over Stages 1 and 2 is because Stage 3 takes into consideration additional important features, which allows the detector to make a more complex classification boundary between the benign and malicious samples. This results in further increase in $DR$, decrease in $FA$, and increase in $HD$. Finally, we can observe from Table \ref{compare2} that the use of multiple data sources improves the $HD$ from $91.83\%$ to $95.66\%$ (i.e., about $4\%$ more increase in the $HD$ compared to using only the net meter readings).

\section{Conclusion} \label{conclusion}
In this paper, detection of false-reading attacks in the net-metering system has been investigated for the first time. Specifically, a new set of attacks tailored for the net-metering system has been proposed to mimic the behavior of malicious customers, and then they have been used to create malicious samples from a dataset of real power consumption and generation readings. Then, data analysis has been performed to detect the time correlations between the net meter readings and the correlations between the readings and relevant data obtained from trustworthy sources such as the solar irradiance and temperature. Based on the data analysis, a general multi-data-source deep learning-based detector has been proposed to identify the false-reading attacks. Our detector has been trained on the net meter readings besides relevant data from trustworthy sources to learn the correlation between them. Extensive experiments have been conducted, and the results indicated that our detector can accurately identify the false-reading attacks. Moreover, the results indicate that our multi-data-source detector achieves higher $DR$ and lower $FA$ than a single-data-source detector trained only on the net meter readings.

\section*{Acknowledgement}
The authors extend their appreciation to the Deputyship for Research \& Innovation, Ministry of Education in Saudi Arabia for funding this research work through the project number 589.

\newcolumntype{P}[1]{>{\centering\arraybackslash}p{#1}}
\bibliographystyle{IEEEtran}
\bibliography{main}

\section*{Biographies}
\begin{IEEEbiography}[{\includegraphics[width=1in,height=1.25in,clip,keepaspectratio]{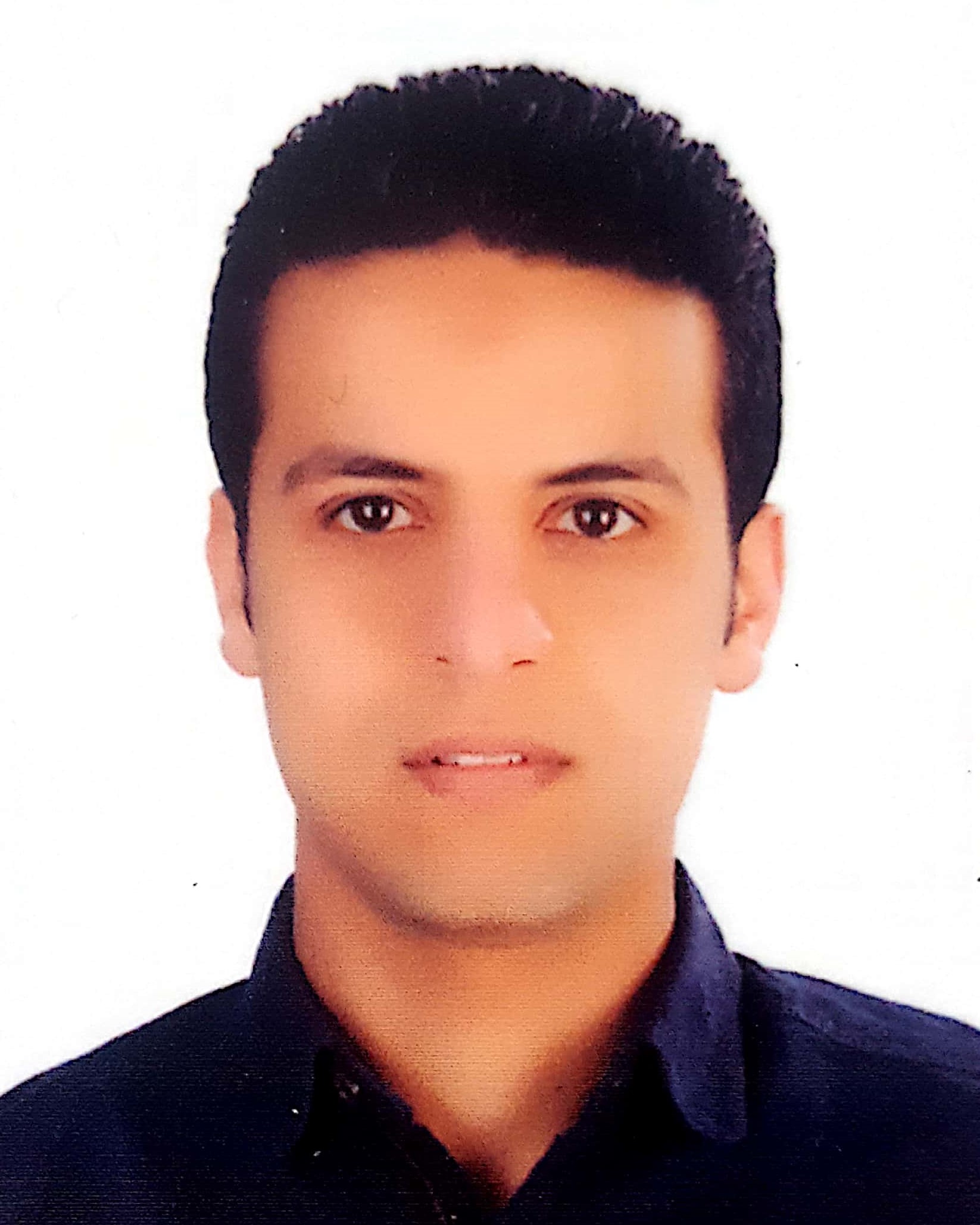}}]{Mahmoud M. Badr}
is currently a Graduate Research Assistant in the Department of Electrical \& Computer Engineering, Tennessee Tech. University, TN, USA and pursuing his Ph.D. degree in the same department. He is also holding the position of a Lecturer Assistant at the Faculty of Engineering at Shoubra, Benha University, Egypt.
He received the B.S. and M.S. degrees in Electrical Engineering from Benha University, Cairo, Egypt in 2013 and 2018, respectively.
His research interests include blockchain, cryptography, 5G networks, network security, and smart grids.
\end{IEEEbiography}
\begin{IEEEbiography}[{\includegraphics[width=1in,height=1.25in,clip,keepaspectratio]{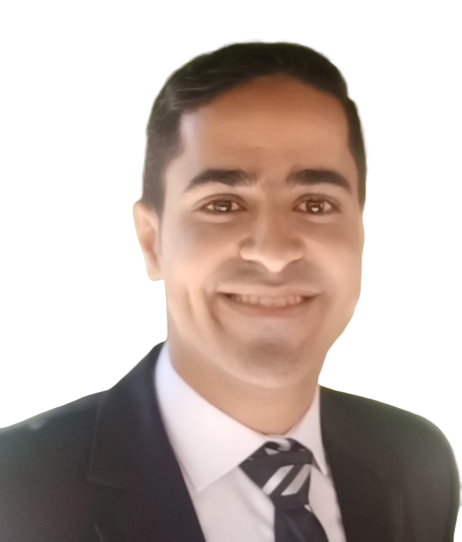}}]{Mohamed I. Ibrahem}
	is currently a Graduate Research Assistant in the Department of Electrical \& Computer Engineering, Tennessee Tech. University, USA and pursuing his Ph.D. degree in the same department.
	He received the B.S. degree and the M.S. degree in Electrical Engineering (electronics and communications) from Benha University, Cairo, Egypt in 2014 and 2018, respectively.
	His research interests include machine learning, cryptography and network security, and privacy preserving schemes for smart grid communication and AMI networks.
\end{IEEEbiography}

\begin{IEEEbiography}[{\includegraphics[width=1in,height=1.25in,clip,keepaspectratio]{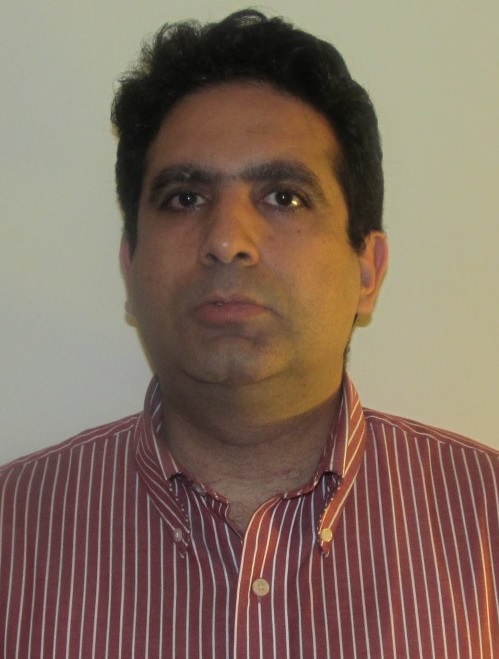}}]{Dr. Mohamed M. E. A. Mahmoud}
	received PhD degree from the University of Waterloo in April 2011. From May 2011 to May 2012, he worked as a postdoctoral fellow in the Broadband Communications Research group - University of Waterloo. From August 2012 to July 2013, he worked as a visiting scholar in University of Waterloo, and a postdoctoral fellow in Ryerson University. Currently, Dr. Mahmoud is an associate professor in Department Electrical and Computer Engineering, Tennessee Tech University, USA. The research interests of Dr. Mahmoud include security and privacy preserving schemes for smart grid communication network, mobile ad hoc network, sensor network, and delay-tolerant network. Dr. Mahmoud has received NSERC-PDF award. He won the Best Paper Award from IEEE International Conference on Communications (ICC'09), Dresden, Germany, 2009. Dr. Mahmoud is the author for more than twenty three papers published in major IEEE conferences and journals, such as INFOCOM conference and IEEE Transactions on Vehicular Technology, Mobile Computing, and Parallel and Distributed Systems. He serves as an Associate Editor in Springer journal of peer-to-peer networking and applications. He served as a technical program committee member for several IEEE conferences and as a reviewer for several journals and conferences such as IEEE Transactions on Vehicular Technology, IEEE Transactions on Parallel and Distributed Systems, and the journal of Peer-to-Peer Networking.
\end{IEEEbiography}
\begin{IEEEbiography}[{\includegraphics[width=1in,height=1.25in,clip,keepaspectratio]{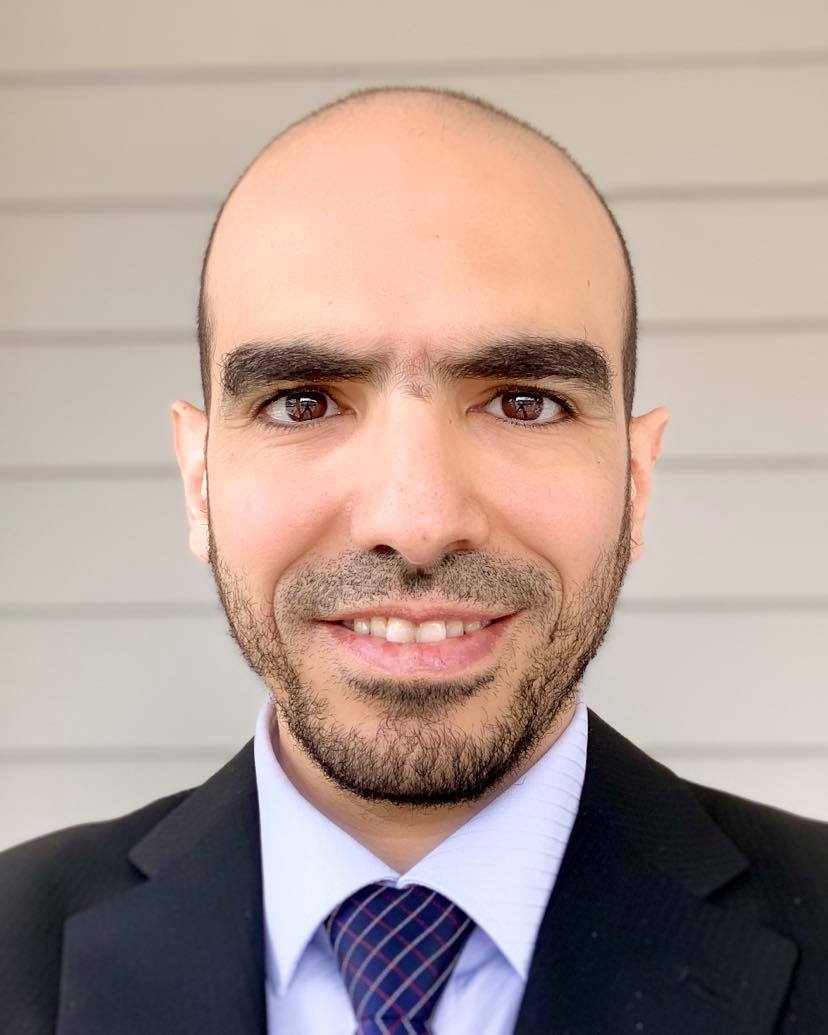}}]{Dr. Mostafa M. Fouda} (SM'14) is currently an Assistant Professor with the Department of Electrical and Computer Engineering, Idaho State University, ID, USA. He also holds the position of Associate Professor at Benha University, Egypt. He has served as an Assistant Professor with Tohoku University, Japan. He was a Postdoctoral Research Associate with Tennessee Technological University, USA. He received his Ph.D. degree in Information Sciences from Tohoku University, Japan in 2011.  His research interests include cyber security, machine learning, blockchain, IoT, 5G networks, smart healthcare, and smart grid communications. He has published over 30 papers in IEEE conference proceedings and journals. He has served on the technical committees of several IEEE conferences. He is also a reviewer in several IEEE Transactions/Magazines and an associate editor of IEEE Access. He is a Senior Member of IEEE.
\end{IEEEbiography}


\begin{IEEEbiography}[{\includegraphics[width=1in,height=1.25in,clip,keepaspectratio]{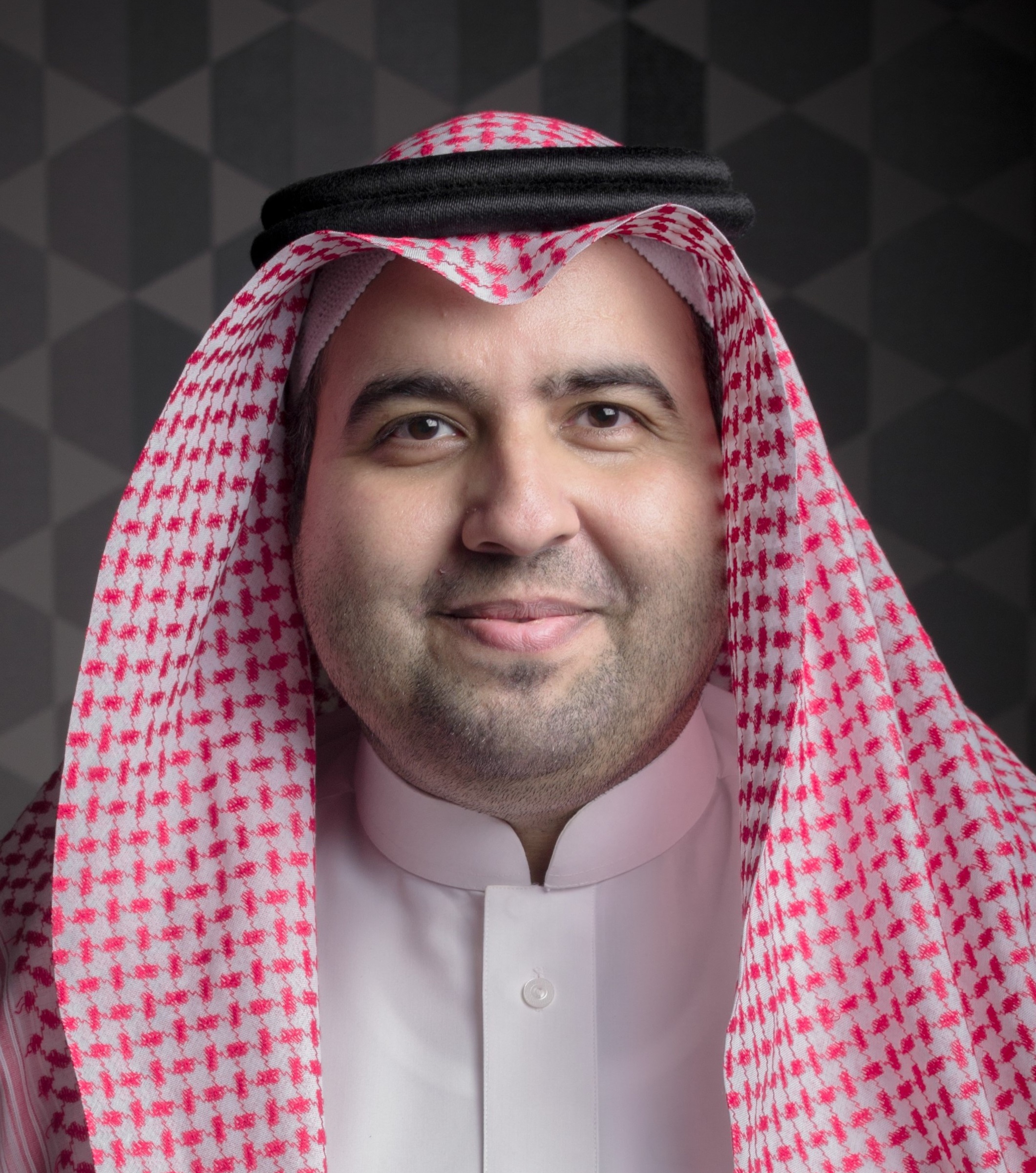}}]{Dr. Waleed Alasmary} (SM'19) received the B.Sc. degree (Hons.) in computer engineering from Umm Al-Qura University, Saudi Arabia, in 2005, the M.A.Sc. degree in electrical and computer engineering from the University of Waterloo, Canada, in 2010, and the Ph.D. degree in electrical and computer engineering from the University of Toronto, Toronto, Canada, in 2015. During his Ph.D. degree, he was a Visiting Research Scholar with Network Research Laboratory, UCLA, in 2014. He was a Fulbright Visiting Scholar with CSAIL Laboratory, MIT, from 2016 to 2017. He subsequently joined the College of Computer and Information Systems, Umm Al-Qura University, as an Assistant Professor of computer engineering, where he currently holds an Associate Professor position. His ``Mobility impact on the IEEE 802.11p'' article is among the most cited Ad Hoc Networks journal articles list from 2016-2018. He is currently an Associate Editor for the Array journal.
\end{IEEEbiography}


\end{document}